\documentclass[useAMS,usenatbib]{mn2e}
\usepackage{graphicx}
\usepackage{subfigure}

\title[On how the optical depth tunes the effects of ISM neutral atom 
flow on debris disks]
{On how the optical depth tunes the effects of ISM on debris disks}
\author[F. Marzari and P. Thebault]{F. Marzari$^{1}$ and P. Thebault$^{2}$\\
$^{1}$Dept. of Physics, University of Padova, 35131 Italy\\
$^{2}$Observatoire de Paris, Meudon, Paris, France}
\begin{document}

\date{Accepted .....;  Received ..... ; in original form ........}

\pagerange{\pageref{firstpage}--\pageref{lastpage}} \pubyear{.....}

\maketitle

\label{firstpage}

\begin{abstract}
The flux of ISM neutral atoms surrounding stars and their
environment affects the motion of dust particles
in debris disks, causing a significant dynamical evolution. 
Large values of eccentricity and
inclination can be excited and strong correlations settle in 
among the orbital angles. 
This dynamical 
behaviour, in particular for bound dust grains, 
can potentially cause significant asymmetries in 
dusty disks around solar type stars which might be detected 
by observations. 
However, the amount of orbital 
changes due to this non--gravitational perturbation is strongly limited by the 
collisional lifetime of dust particles. We show that for large values of the 
disk's optical depth the influence of ISM flow on the disk shape is almost
negligible because the grains are collisionally destroyed before they can 
accumulate 
enough orbital changes due to the ISM perturbations. 
On the other hand, for values smaller than $10^{-3}$, peculiar asymmetric patterns
appear in the density profile of the disk when we consider 
1-10 mum grains, just above 
the blow-out threshold.
The extent and relevance of these asymmetries grow for lower
values of the optical depth. An additional 
sink mechanism, which may prevent the formation of 
large clumps and warping in the disks
is related to the fast inward migration due to the drag
component of the forces. When a significant eccentricity is 
pumped up by the ISM perturbations, the drag forces 
(Poynting-Robertson and in particular ISM drag)
drive the disk particles
on fast migrating tracks leading them into the 
star on a short timescale. It is then expected that disks 
with small optical depth expand inside the parent body ring
all the way towards
the star while disks with large optical depth would not 
significantly extend inside. 

\end{abstract}


\section{Introduction}

The observable part of debris disks are small ($\leq 1\,$mm) dusty or icy grains, collisionally produced from larger, undetectable parent bodies. In addition to the gravitational pull of the star, these grains are also affected by several forces such as stellar radiation pressure, Poynting-Robertson (PR) drag and the possible gravitational influence of large bodies in the neighborhood. As has been shown in numerous numerical studies, the combined effect of these different forces can lead to complex spatial structures in resolved disks \citep[e.g.][]{wyatt08}. 
A less investigated additional force that could have an influence on grain dynamics is the drag due to particles from the surrounding interstellar medium (ISM). The effect of ISM has first been addressed by \citet{arty97}, who studied the level of disk erosion due to sandblasting by ISM dust grains. They concluded that, at least around massive stars, this effect was negligible because small ISM grains felt a strong repulsive radiation force.
More recently,\citet{scherer}, \citet{debes09}, \citet{manes09}, \citet{bera} and 
\citet{pasto} considered instead the effect of ISM {\it neutral atoms} on disk grains. This flux of neutral atoms acts indeed similarly to the solar wind or radiation pressure from a physical point of view but, being monodirectional, can significantly perturb the trajectories of the grains, and potentially induce asymmetric structures in the disk. In particular  \citet{manes09} and \citet{debes09} suggest that the ISM flux can explain the unusual morphology of some debris disks like HD61005 and  HD32997. 
In their model \citet{debes09} consider dust particles close to the blow-out size for the star
and compute the trajectories of perturbed grains over a timescale of 5000 yrs. 
The majority of their
grains are strongly perturbed and end up quickly on hyperbolic orbits. A similar scenario is 
outlined by  \citet{manes09} where they concentrate on small grains (0.1 $\mu$m) 
whose lifetime before ejection is of the order of a few $10^3$ years. The morphology changes they observe is mostly due to the fast transfer of grains from low eccentricity orbits into 
hyperbolic trajectories. 

In this paper  we take these studies a step further and we concentrate on the
effects of the ISM neutral flow on bound Keplerian orbits of dust particles in
debris disks around solar type stars. 
The orbital changes on relatively large dust grains (we model grains with 
radius ranging from 1 to 10 $\mu$m) is slow and it occurs on timescales of the order of 
$10^6$ yrs. This timescale is much longer than that considered in 
\citet{debes09} and \citet{manes09} and it is related to the larger size
of the grains. In addition, our grains are bound to the star 
and the orbital evolution induced by the ISM flow lead  
to eccentric and possibly asymmetric disks,
characterized by density clumps, only if the collisional lifetime of the 
grains allows it.
Thus, a crucial parameter that controls the efficiency of the ISM flow 
in shaping debris disks is their geometrical optical depth, on which 
collisional lifetimes directly depend. The collisional cascade that 
is steadily eroding, by cratering and fragmentation, all solid bodies 
in a debris disk, does indeed reduce the lifetime of dust grains, thus limiting the amount
of time their 
trajectories can be perturbed by non--gravitational forces like 
radiation pressure, PR drag and interaction with ISM \citep{wyatt05}. 
To correctly evaluate the 
impact of ISM on the density profile of a dust disk we  thus need to account for
the limited lifetime of individual particles and the amount of orbital changes 
that they can accumulate 
during that time. 

The role of collisions had
been discarded by \citet{debes09} and \citet{manes09} because their 
dust grains are pushed on hyperbolic trajectories by the ISM flow on a 
very short timescale (a few $10^3$ years), comparable to the collisional lifetime
of the particles they handle. 
In our scenario, grains remain on bound orbits but a simple numerical 
integration of dust 
trajectories over a long timespan, performed to predict the overall disk 
density distribution, can lead to misleading result which 
are correct only for disks with very low optical depth.
A numerical model intended to evaluate the impact of ISM on 
the morphology of dense disks must include an estimate of the lifetime 
of each individual particle that will be part of the disk 
only for an interval of time no longer than its collisional lifetime
after which it will be replaced by a new one. If the integration time of the 
whole particle ensamble is a few times longer 
than the lifetime of the considered dust particles, we will obtain
a stationary relaxed population of grains which will describe 
in a reliable way the spatial distribution of the debris disk.
We must also account that dust of different sizes would have different
lifetimes so the presence of density structures would depend on the size of the
dust that is imaged. 

The effect of ISM neutral atom flux on
the dust grain trajectories is twofold. On one side it forces a coupled evolution of 
eccentricity and pericenter longitude that drives the particles on aligned 
eccentric orbits. On the other side, if the ISM flow is inclined with respect
to the parent body orbital plane a significant inclination can be 
pumped up. In addition, a fast inward migration is forced by the 
PR drag, which is strong for highly eccentric orbit, and by the 
ISM drag component of the force related to difference between the 
monodirectional velocity of the ISM neutral atoms and the orbital 
velocity of the dust particles. When they migrate close to the 
star, the grains either sublimate or impact potential planets or 
the star itself going lost. 
This is an additional sink mechanism, purely 
dynamical, which contributes  
to the erosion of the dust population. 

In this paper we intend to address these issues concentrating on the 
balancing between the orbital evolution of dust grains in dusty disks
under the effect of ISM neutral atoms perturbations and collisional 
lifetime. Our goal is to outline the features of the steady state population
of dust produced by the combination of these effects. We will explore the 
amount of asymmetry in the debris disk created by the ISM perturbations 
and how this depends on the optical depth of the disk. 
We concentrate in particular on dust grains of
1 um size since these particles, which are just above the cut-off size
imposed by radiation pressure($\sim 0.6$ $\mu$m), are those that dominate
the geometrical cross section of the system, and thus the flux coming
from the disk in the visible and near-IR (see for example
\cite{theau}). We will also give a hint to the evolution of 10 um size
particles to test how the density distribution of the disks depends on 
the observed particle size. 

\section[]{The numerical model}

Our numerical integration scheme computes the trajectories of a large number 
of dust particles of a given size using RADAU \citep{rad}. It is a 
very precise algorithm that accurately handles highly eccentric orbits as those 
produced by ISM perturbations.  It is an implicit Runge--Kutta integrator 
of 15th-order which
proceeds by sequences within which the substeps are taken at Gauss-Radau spacings. 
High orders of accuracy are obtained with relatively few function evaluations. In addition, 
after the first sequence, the information from previous sequences is used to improve the accuracy
and the integrator itself chooses the next sequence size depending on the estimated 
variability of the 'force' term in the differential equations.  
Non--gravitational forces like 
radiation pressure and PR drag are included as in
\cite{marva}  and, in addition, we 
model the effects of the ISM neutral gas atoms adopting the same approach 
used to calculate the solar wind pressure (\citet{scherer}). The acceleration 
of a dust particle due to the 
impact with fast neutral atoms of the ISM wind is computed as:

\begin{equation}
{\bf f} = - C_D (n_H m_H) A_s \mid {\bf v} - {\bf v_{H}}\mid ({\bf v} - {\bf v_{H}})
\end{equation} 

where $v$ is the orbital velocity of the dust grain and $v_H$ is the 
velocity of the ISM neutral atoms.  
$m_H$ and $n_H$ are the mass and density of neutral hydrogen atoms,
 respectively, $A_s = \pi r^2$ is the area of the dust particle of 
radius $r$ while $C_D$ is a drag coefficient whose value is about 2.5.
This functional dependence is robust only in the limit where 
the gas of the ISM has a temperature lower than about 5000 K. Under this condition, 
the thermal speed is lower than $v_H$ and the momentum change is 
proportional to the $(v - v_{H})^2$. For hot ($T > 5000 K$) ISM clouds the momentum change 
would be proportional to $(v - v_{H})$ \citep{bera,drsal}.

For the collisional lifetime of each particle, we adopt the usual expression

\begin{equation}
t_c \sim {{T} \over  {12 \tau}}
\end{equation}

where $\tau$ is the local geometrical vertical optical depth of the disk and $T$
the orbital period of massless particles at this location \citep[e.g.][]{arty97}.
Although detailed numerical studies (e.g., \cite{theau}) show that real collision rates 
may strongly depart from this simplified expression, we adopt it here for its simplicity and because it is an easy way to tune in the collisional effects.

Particle size is a crucial parameter since it determines the response to radiation pressure 
parameterized by $\beta$, the ratio of radiation pressure to stellar gravity. 
Given the steep size distribution for collisional cascades and the limited number of test 
particles in a deterministic code, it is impossible to consider a wide size range. 
We thus restrict our study to several independent simulations each with single size particles. 
These different runs can then be combined, when properly weighted, to give a 
first-order estimate of ''real'' disks.
 We consider a scenario where the central body is a solar type star and  
the dust grains of the debris disk are the by--product of collisions 
occurring in a ring of parent bodies 
(planetesimals, asteroids, comets) on circular orbits. 
Whenever a dust particle is produced, 
its orbital elements are computed from the position and velocity vectors of 
the parent body accounting for the reduced gravity $1 - \beta$ 
due 
to radiation pressure. The average collisional lifetime $t_c$ is derived for 
the particle and, at starting, a random value of 'age' $t_a$ is given 
to each grain. As the integration advances, $t_a$ is updated with the 
timestep $\Delta t$. When 
$t_a$ becomes larger than the collisional lifetime $t_c$, the particle
is eliminated from the sample and a new one is generated with new 
orbital elements and $t_a = 0$. When the timespan of the integration is 
a few times longer than $t_c$ we have a steady state population of 
dust particles where the collisional lifetime determines for how long
a grain is subject to non--gravitational perturbations.  In addition, when a particle 
is injected on a hyperbolic orbit or it has migrated far inside the parent body ring
it is discarded and a new one is 
generated. As a consequence, the steady state population can be due to a balance
between the different mechanisms of dust elimination i.e. collisions,  
ejection out of the system or inwards migration. 
We assume in our model 
that the parent bodies are large enough not to be significantly affected
by ISM. This is a reasonable assumption since the forces acting on the 
particles strongly depend on their size. Larger parent bodies would be
affected by ISM only on a much longer timescale.

\section{Initial setup}

We consider as a test case for our modeling a disk produced by a ring of parent 
bodies evenly distributed in between
50-70 AU on circular orbits. 
Since the size of the ring is small, assuming a uniform
distribution is a good approximation. We adopt different values for the
optical depth of the disk ranging from $\tau \sim 10^{-3}$, typical of
dense collision--dominated disks like that around Beta Pictoris, to $\tau \sim 10^{-6}$ for
almost collisionless disks similar to the Kuiper Belt where 
transport mechanisms like PR drag, stellar or ISM winds can dominate.  
The ISM gas flow is
approximated as a monodirectional flux moving along the x-axis with a
velocity of 20 km/s and a concentration of neutral hydrogen atoms equal to 
$n_H = 0.1$ $cm^{-3}$. These are typical values of the interstellar
cloud surrounding our solar system. The local speed of the sun 
with respect to the
ISM is around $25 \pm 2$ $km/s$ but some deceleration of the
interstellar neutral hydrogen flow across the heliospheric interface is
expected reducing the speed to about $22$ $km/s$ \citep{lalle}. The density
we adopt in the numerical model is intermediate between the value given
by \cite{fahr96} equal to$n_H = 0.05$ $cm^{-3}$ and that given in \cite{fri99}
of $n_H = 0.22$ $cm^{-3}$. 

\section{The coplanar case: ($ e, \varpi$) evolution}

\begin{figure*}
\begin{center}
\begin{tabular}{c c}
 \hskip -3 truecm
 \resizebox{120mm}{!}{\includegraphics[angle=-90]{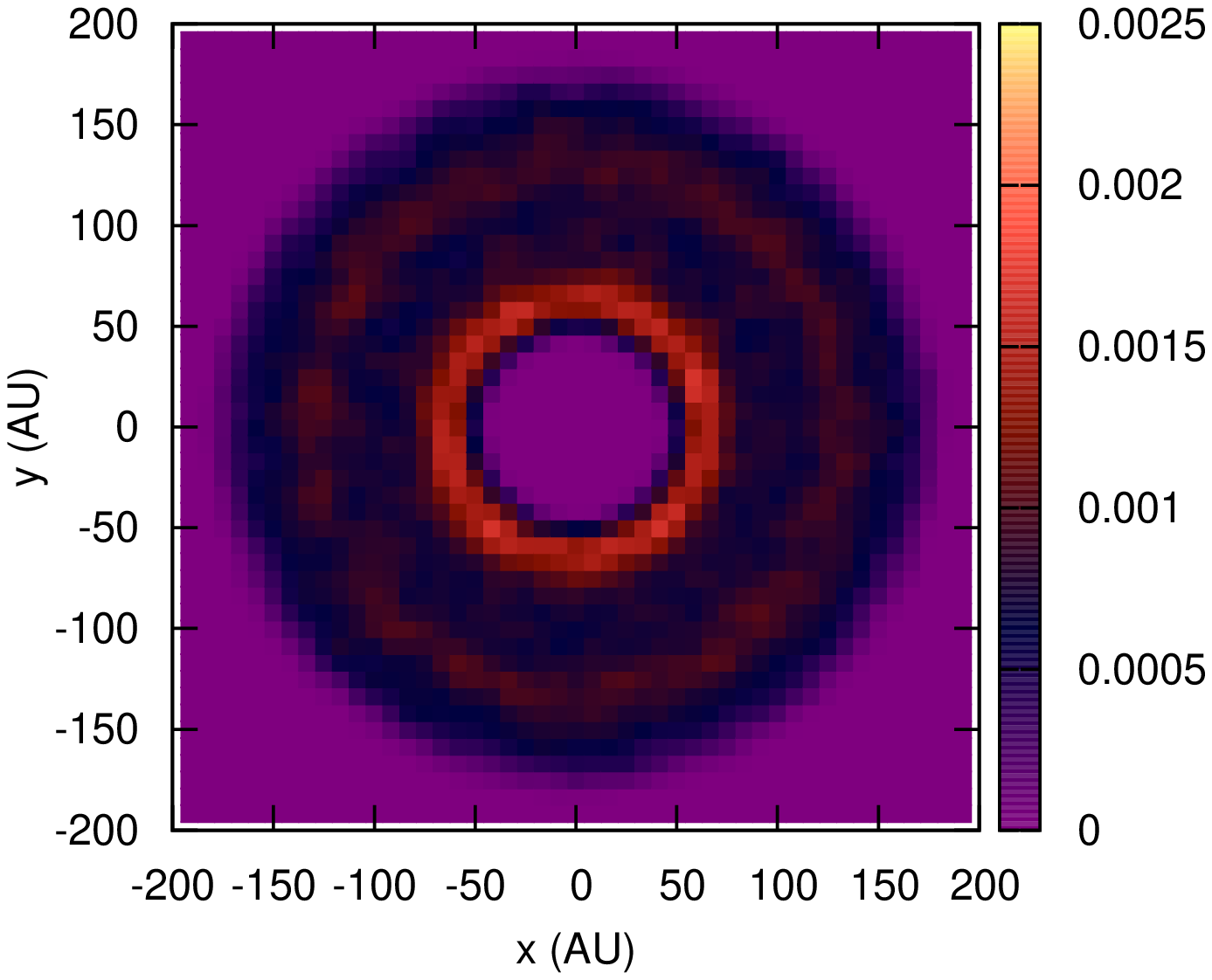}}
 \hskip -3 truecm
 \resizebox{120mm}{!}{\includegraphics[angle=-90]{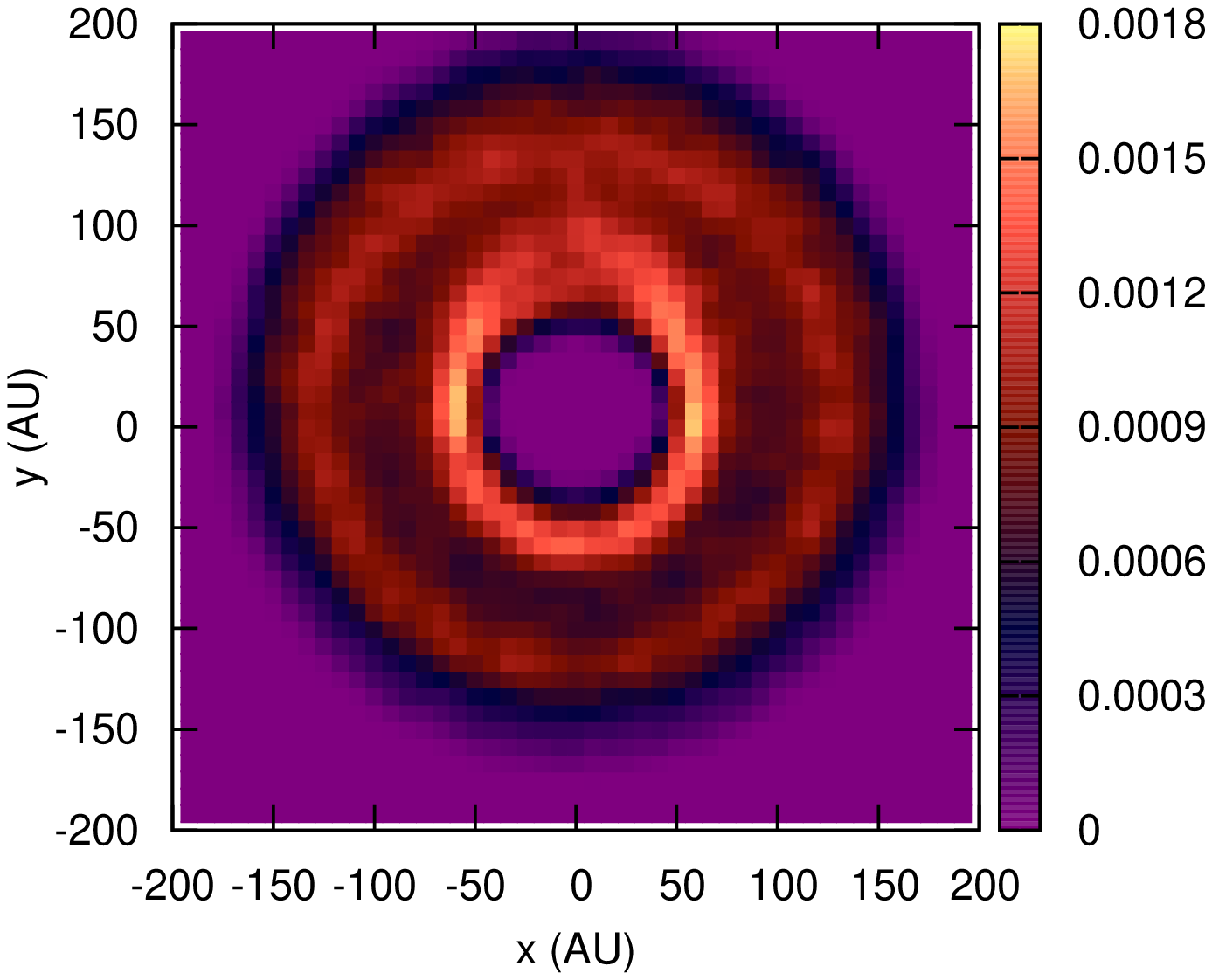}}\\
 \hskip -3 truecm
 \resizebox{120mm}{!}{\includegraphics[angle=-90]{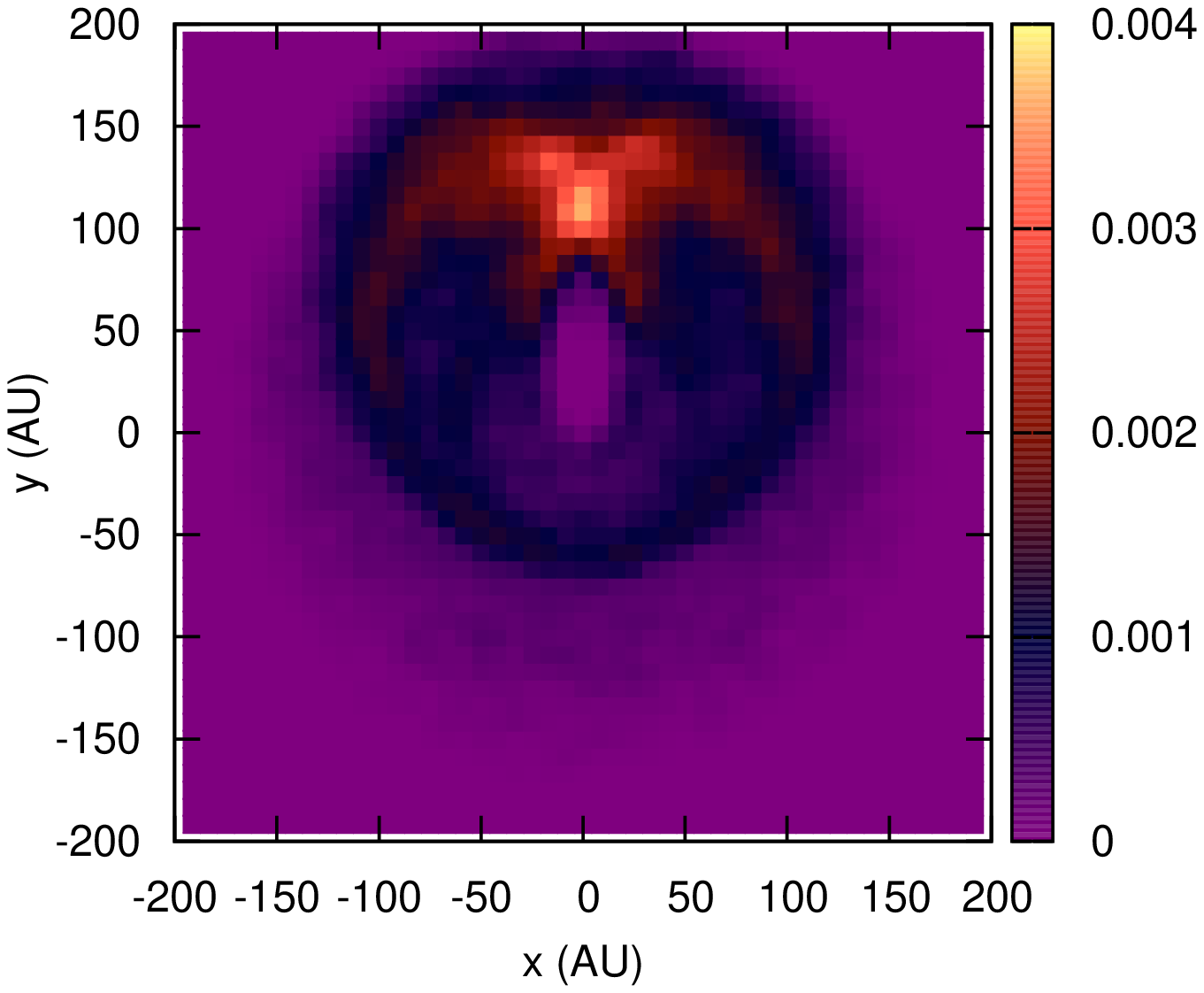}} 
 \hskip -3 truecm
 \resizebox{120mm}{!}{\includegraphics[angle=-90]{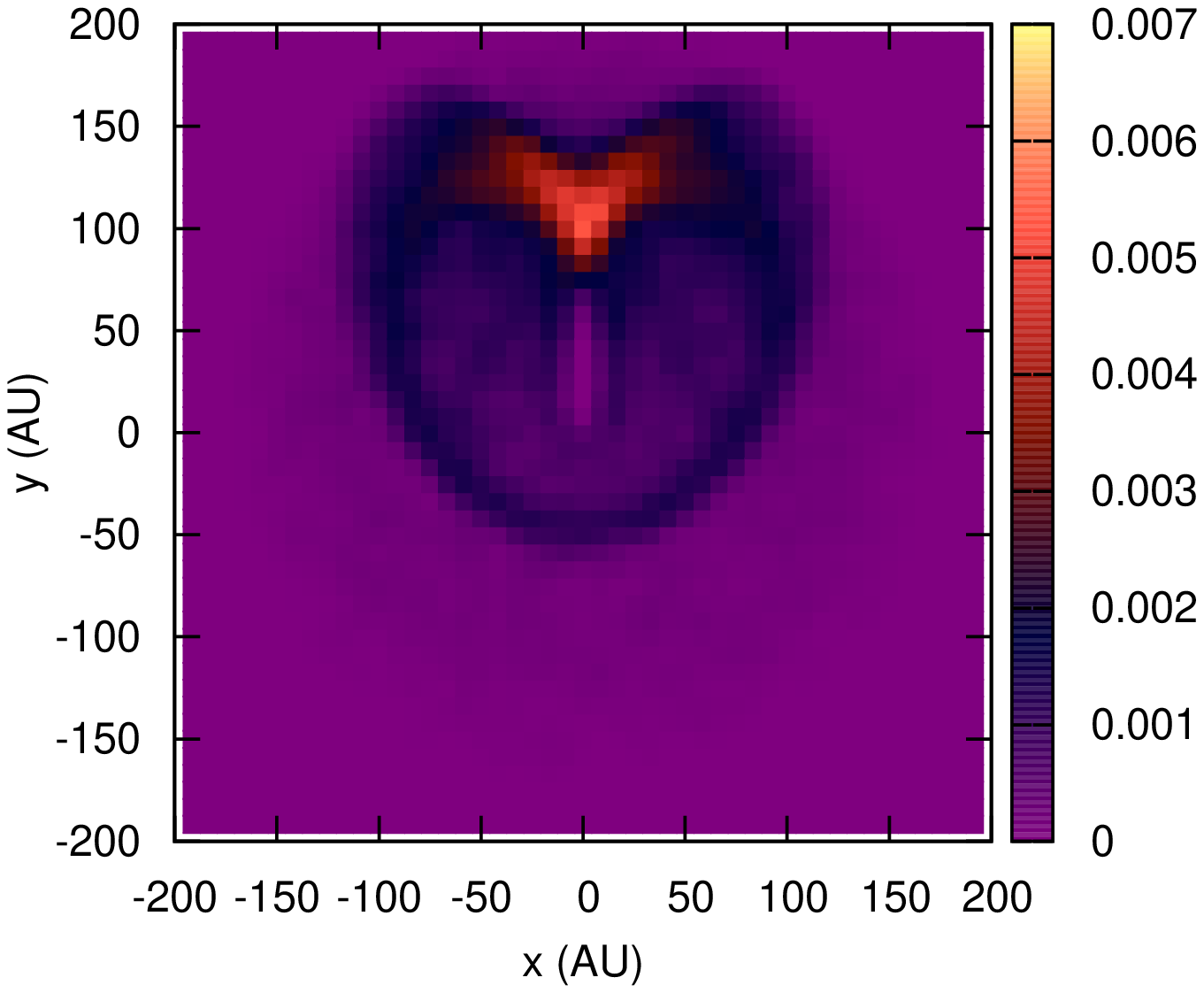}} 
 \end{tabular}
 \caption{2--D density distribution of 1 $\mu$m size dust particles in debris disks 
generated by a ring of parent bodies moving on circular orbits in between 
50--70 AU. 
The density is computed as number of particles populating a squared region 
at a given timestep 
divided by the total number of particles in the model.
The top--left
plot shows how the disk appears when the ISM flow on the grains is neglected. The 
top--right plot illustrates the density distribution when ISM is acting on the 
particles and the optical depth is $\tau = 1 \times 10^{-3}$. The two bottom plots 
show the density distribution when $\tau = 1 \times 10^{-4}$ and $\tau = 1 \times 10^{-6}$,
respectively.}
\label{f1}
\end{center}
\end{figure*}

We first assume that the interstellar gas flow lies on the 
plane of the parent body ring. This may be considered 
a test bench  where to study the evolution of the 
($ e, \varpi$) orbital elements. In effect, it may not be 
a realistic scenario since the 
flow, impinging on one side of the disk, may not reach with 
equally intensity the other side being eventually absorbed.
This does not occur when there is some inclination between the dust
orbits and the ISM flow. 
However, it is useful to explore the density distribution in 
the planar case since the structures and clumpings that develop in the $(x--y)$ plane
are retrieved when  a low inclination between the flux and the initial 
dust orbits is assumed. 
In Fig.\ref{f1} the position of 1 $\mu m$ size  particles
are given after an evolution lasting about
3 times the average collisional lifetime for the case with
$\tau = 1 \times 10^{-3}$ ($\Delta t = 0.5$  Myr)  and 
$\tau = 1 \times 10^{-4}$ ($\Delta t = 5$  Myr, about twice $T_{stark}$ 
defined by formula (7)). In the case 
with $\tau = 1 \times 10^{-6}$ the collisional timespan is around 
$100$ Myr, however the fast inward evolution due to the large 
eccentricity, achieved by the dust grains because of the 
ISM perturbations, significantly limits the lifetime of the particle 
within the disk. 
The most relevant 
sink mechanism in this scenario is related to the fast inward migration
of grains driven by both the 
PR drag and the ISM drag components of the force. 

The density plots are derived by computing the number of particles 
populating the local spatial area (equal size squared 
bins) at the end of the numerical simulation. This number is then normalized to the 
total number of particles which is kept constant and equal to 
$4 \times 10^4$ in all our simulations. 

The first plot on the top left in Fig.\ref{f1} shows how the debris disk 
appears
when the ISM force is neglected. It is axis--symmetric with 
a density peak at the internal ring corresponding to the location 
of the parent bodies. This is where all the pericenters of the 
particles reside  and the higher density is due both to the 
geometrical configuration (the orbit is tangent to the 
circle) and to the creation of new grains when the older ones
are destroyed. A second  peak in the density plot appears at the outer ring
where the apocenters of the particles are located and 
the grains move at the lowest speed. 
The pericenter longitudes are randomly 
distributed in between 0 and $360^o$ when the steady state is reached. 
The initial eccentricities of our sample of
particles
are encompassed
between  0.42--0.45 while the initial semimajor axes range from 90 to 120 AU.  
This orbital distribution is due to the value of $\beta$ adopted 
at the beginning of the simulation that reduces the mass of the 
central star injecting the grains into eccentric orbits. 

When the ISM effects are included, asymmetries develop in the disk due to the 
eccentricity and pericenter evolution forced by the ISM perturbation. 
These asymmetries are barely detectable when $\tau = 10^{-3}$ (Fig.\ref{f1},
upper right plot) but they
become more pronounced for lower values of $\tau$ (Fig.\ref{f1}, bottom plots)
 since the 
grains have more time to accumulate the perturbative effects. 
When $\tau = 10^{-3}$ the two peaks at the particles pericenters and apocenters 
are still visible but the disk begins to develop an eccentric shape. 
The alteration in the disk shape is more noteworthy for 
$\tau = 10^{-6}$ when the disk appears as the superposition 
of two ellipses. 

\begin{figure}
 \hskip -2.4 truecm
 \resizebox{120mm}{!}{\includegraphics[angle=-90]{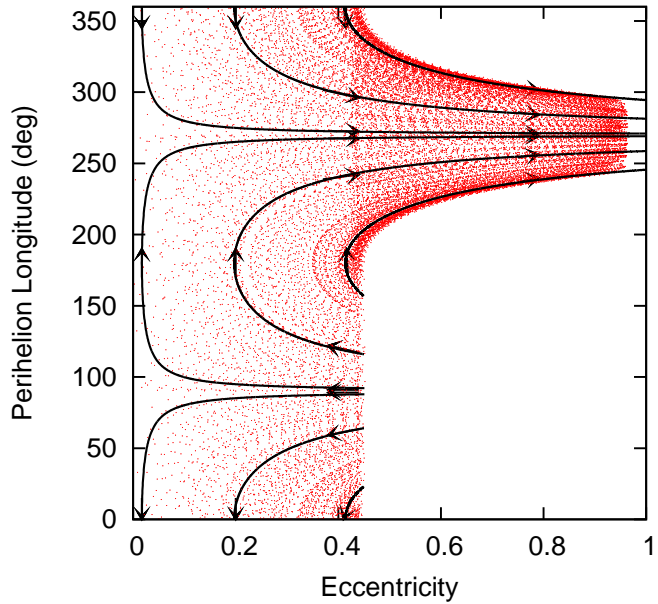}} 
 \caption{Evolution in the ($e,\varpi$) plane of 100 sampled dust 
particles. The orbital elements are computed over an interval of time of 
5 Myr. 
The optical depth $\tau$ is 
set to $10^{-6}$ to grant enough time to the grains to dynamically evolve.  
}
\label{f2}
\end{figure}

This feature can be explained by inspecting the orbital behaviour 
of the grains under the effect of the ISM neutral gas. In Fig.\ref{f2}
we show the evolution with time of the grain orbits in the 
$(e,\varpi)$ plane. 
Particles evolve towards high eccentricity values ($\sim 1$) while, at 
the same time, the pericenter values tend to $270^{\circ}$.
When large values of eccentricity are achieved, the particles 
drift quickly inside because of the drag component of the forces. 

This behaviour can be interpreted on the basis of the analytical solution of the
classical Stark problem \citep{bera,pasto} where the Keplerian motion of 
the particle is perturbed by an external force constant in magnitude and direction. 
In effect, the dust particle dynamics is different from that predicted by the 
classical Stark problem since it 
includes PR drag which is a tangential force. 
In addition,  the ISM force 
depends on the difference between the particle $v$ and the neutral atom velocity 
$v_H$ so that the force changes periodically depending on the position of the particle in its
orbit. In our scenario, $v$ is approximately 5--10\% of $v_H$ but it 
can become as large
as 30\% for highly eccentric orbits once close to perihelion.
This has a significant effect on the semimajor axis of the particle 
orbit causing a fast inward migration. 
However, the analytical equations that describe the evolution of 
a body in the general Stark problem gives a very accurate description of the 
evolution of eccentricity, pericenter longitude, inclination and nodal longitude. 
The secular solution, obtained by 
\cite{bera} after averaging the Hamiltonian of the system (see their Section 3.2), 
is briefly summarized hereinafter since it will help in interpreting 
the numerical results. The authors start their analytical development by
 aligning the constant force 
along the z--axis and find that 
the system admits 3 integrals of motion
\begin{eqnarray}
 K_z & = & \sqrt{(1 - e^2)} \mid cos i  \mid \\
 I_H & = & (1 - K^2)(1 - K_z^2/K^2) sin^2 \omega 
\end{eqnarray}
and the semimajor axis, 
where $K(t) = \sqrt(1 - e(t)^2)$ is the dimensionless total angular momentum 
which is not conserved, while $K_z$ is conserved since the force is 
parallel to the z--axis. By solving the Hamilton equations, they find that 
$K$ oscillates between a maximum and minimum value given by
\begin{eqnarray}
 K^2_{M,m} = \frac {1 + K_z^2 -I_H^2}{2} \pm 
\left [\left( \frac {1 + K_z^2 -I_H^2}{2} \right)^2 - K_z^2 \right ]^{1/2}
\end{eqnarray}
The maximum and minimum values of the angular momentum translates into minimum
and maximum values of eccentricity which depend on the 
inclination of the ISM flux respect to the initial orbit. 
The time dependence of $K$ is 
periodic and
given by 

\begin{eqnarray}
 K(t) & = & \frac{K^2_m + K^2_M}{2} + \frac{K^2_M - K^2_m}{2} 
sin \left(4 \pi \frac{t -t _0}{T_{stark}} \right)
\end{eqnarray}

The period of the cycle $T_{stark}$ is 

\begin{eqnarray}
T_{stark} = \frac{4 \pi m_g}{3 S} \sqrt{\frac{G M_{\odot}}{a}}
\end{eqnarray}

where $a$ is the orbital semimajor axis, $G$ the gravitational constant,  
$M_{\odot}$ the star mass and $S$ is the constant force applied in the 
z--axis direction. $T_{stark}$ is constant in the general Stark 
problem, but it will change with time when we deal with dust particles under the action of 
radiation and ISM forces because of the inward drift that reduces the 
semimajor axis $a$ and tends to circularize orbits. 
As a reference, the average value of $T_{stark}$ for the initial sample of 
 1 $\mu$m particles in our model is about 2.3 Myr. This means that 
the dust grains will achieve an eccentricity of about $1$ on a timescale 
shorter than $T_{stark}/2 \sim 1.1 Myr$. When their eccentricity is close to 
$1$ they drift quickly inside. 

Once derived $K$ as a function of t, we can easily compute 
$e(t)$, $\omega(t)$ (from the constant $I_H$) and $i(t)$ (from $H_z$). The 
solution is closed by the equation that gives the node longitude as 
a function of $K(t)$
\begin{eqnarray}
\Omega (K) = arctan \left( \sqrt{ \frac{ (K^2 - K_m^2)(K_M^2 - K_z^2)}
{K_M^2 - K^2) (K_m^2 - K_z^2)}} \right)
\end{eqnarray}

Thanks to the secular integrability of the Stark problem we can interpret the 
evolution shown in  Fig.\ref{f2} and the density plots in Fig.\ref{f1}.
In the planar case, the maximum value of eccentricity $e_M = 1$ and the dust 
particles follow the evolution predicted by the 
equations of the \cite{bera} model in the $(e,\omega)$ plane. 
However, to properly compare the analytical solution of the Stark problem
with our numerical results, we have to take into account 
the difference in the definition of inclination. The rule we adopt 
in this paper is 
that the inclination is counted from the orbital plane by 
a counter--clockwise rotation around the x--axis. As
a consequence, $i = 0^o$ corresponds to $i = 90^o$ 
in the \cite{bera} model. This introduces a $180^o$ 
difference in the definition of the
perihelion longitude between our numerical solutions and the
analytical formalism of  
\cite{bera}.
In Fig.\ref{f2} the analytical curves for 3 different 
initial conditions in $e$ and $\omega$ are shown as continuos lines 
with arrows pointing in the direction of the time flow. The
particles move from an eccentricity of about 0.4 and, following the 
countours of constant $H_z$ and $I_H$, evolve towards 
$e = 1$ while the pericenter tends towards the asymptotic value 
of $270^{\circ}$. 

This behaviour explains why the shape of the disk in
the density plots, for low 
values of $\tau$, appears as a superposition of two 
ellipses. Most particles cluster either from below or from above 
around the fixed pericenter value of $270^o$. This alignment  
was also predicted by \citet{scherer} and observed 
in \citet{manes09}. Thanks to the analytical solution of \cite{bera} it is possible to 
predict the angle between the two elliptical shapes that stand out in 
Fig.\ref{f1}. The analytical curves that enclose, for large eccentricities,
all the particles, give for $e=1$ a value of $\omega \sim 244.5^o$ and $\omega \sim 295.5^o$. 
Since most of the particles cluster close to these two curves, 
the angular separation between the two ellipses of Fig.\ref{f1} is 
approximately $51^o$. A shortcut to obtain the angular separation 
is to use directly the formula (33) of 
\cite{bera} that, in the planar case, links eccentricity and perihelion to their constant 
initial values through the equation  $ e cos \omega = \Lambda$ where $\Lambda$ is the value of 
$ e cos \omega$ at t=0. By checking Fig.\ref{f2}, we notice that the curves leading to 
the outer edges of the peak in the $(e,\omega)$ plane approximately evolve from  
the following points: $(0.43,180^o)$ and $(0.43,360^o)$. Using these as 
starting values to compute  $\Lambda$, we get $\Lambda = \pm 0.43$. 
As a consequence, the asymptotic values of $\omega$, when $e$ approaches 1, are
$\omega \sim arcos (0.43) \sim 64.5^o + 180^o = 244.5^o$ and 
$\omega \sim rcos (-0.43) \sim  295.5^o  + 180^o = 295.5^o$. The 
same value of separation of $\sim 51^o$ is obtained. 

When the eccentricity becomes large, particles begin to drift inwards 
towards the star at
increasing speed because of the strong PR drag and 
ISM perturbations on the semimajor axis. 
There is a larger variation in the difference between the orbital 
and ISM velocity over an orbit, and this causes an increasingly 
faster inward drift of grains.
This explains
why the size of the disk, for small optical depths, is smaller even if more 
asymmetric. 

In our model, when the particles have a semimajor axis smaller than 20 AU 
they are removed from the representative sample and a new particle is drawn. 
This limit is related to the large eccentricity of these particles 
when they have drifted inside. With a semimajor axis of 20 AU and 
an eccentricity of 0.9 they are very close to the star.  It
is expected that in this situation the dust particles may either 
sublimate or be scattered by internal planets. According to \cite{koba, muka}
micron--sized dust grains consisting of a silicate core with an icy 
mantle may undergo 
active sublimation at temperatures higher than 110 K. This temperature is 
estimated to be present on the surface of a 1 $\mu$m grain at about 
20 AU from a solar type star.  
This scenario, however,  does not easily apply to 
dust particles perturbed by ISM. They are on highly eccentric orbits
while \cite{koba} consider grains on circular orbits
and perturbed by PR drag only. In our model, the drift timescale 
is very short since, in addition to PR drag, a strong
contribution to the drag force is given by the ISM perturbations
and we expect that 
the particle have a rapid infall towards the star when an eccentricity 
larger than 0.9 is achieved. It appears a complex 
issue to predict the amount of sublimation in this context and 
compare it to the drift timescale. For simplicity, we  
assume that when the semimajor axis of a dust particle
crosses the lower limit of 20 AU a series of mechanisms would make its 
contribution to the outer disk negligible. A test run where this limit was 
lowered to 15 AU did not change the outcome in terms of density plots.
Scattering by planets appear also a reasonable outcome when particles 
move close to the star. The presence of a debris disk is a strong
indication that planetary formation occurred in the system. It is then 
expected that a planetary system orbits the star at some distance 
from it. The potential planets would intercept the dust particles as they get 
close to the star eliminating them from the disk population. 

\begin{figure*}
\begin{center}
\begin{tabular}{c c}
 \hskip -3 truecm
 \resizebox{120mm}{!}{\includegraphics[angle=-90]{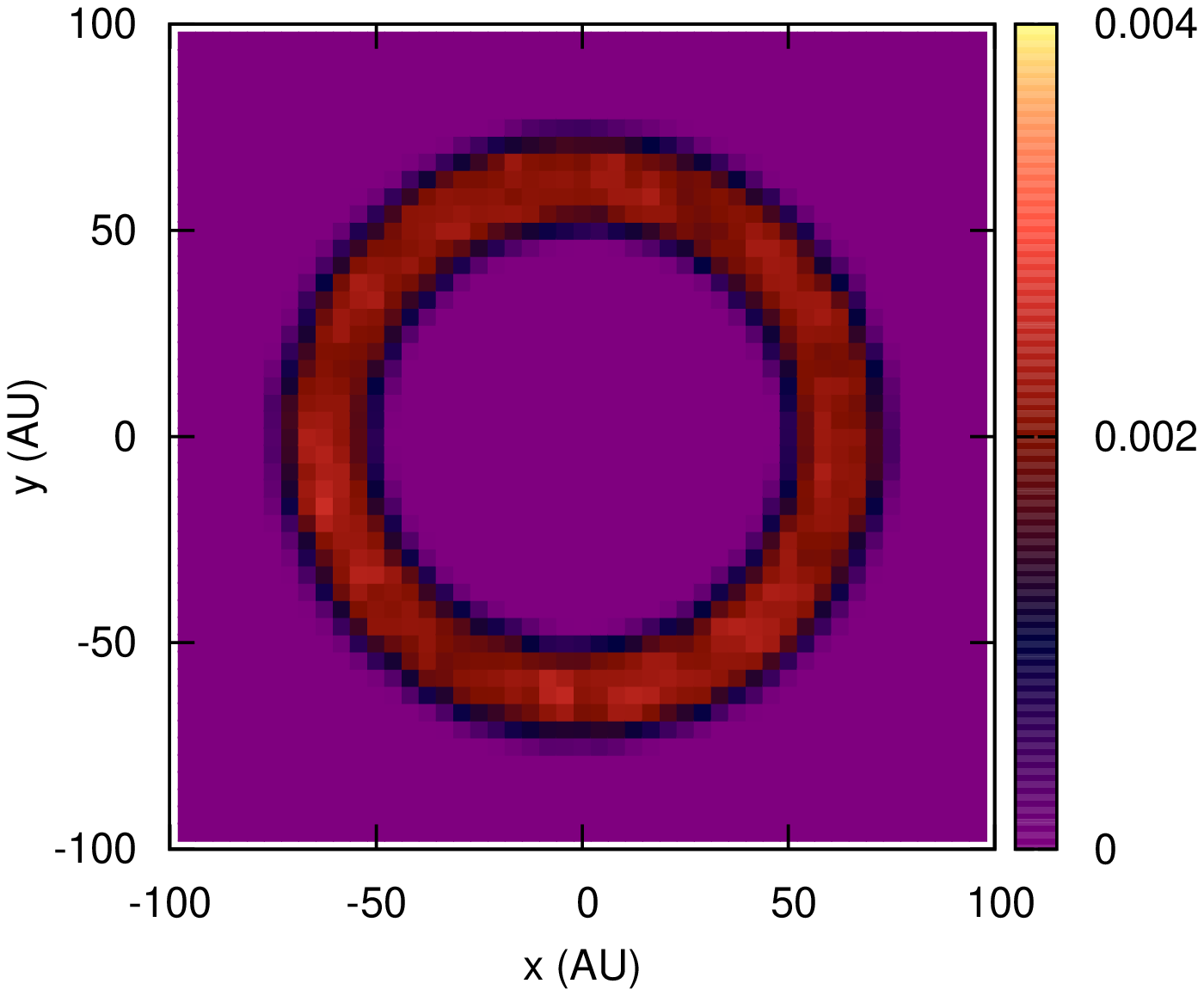}}
 \hskip -3 truecm
 \resizebox{120mm}{!}{\includegraphics[angle=-90]{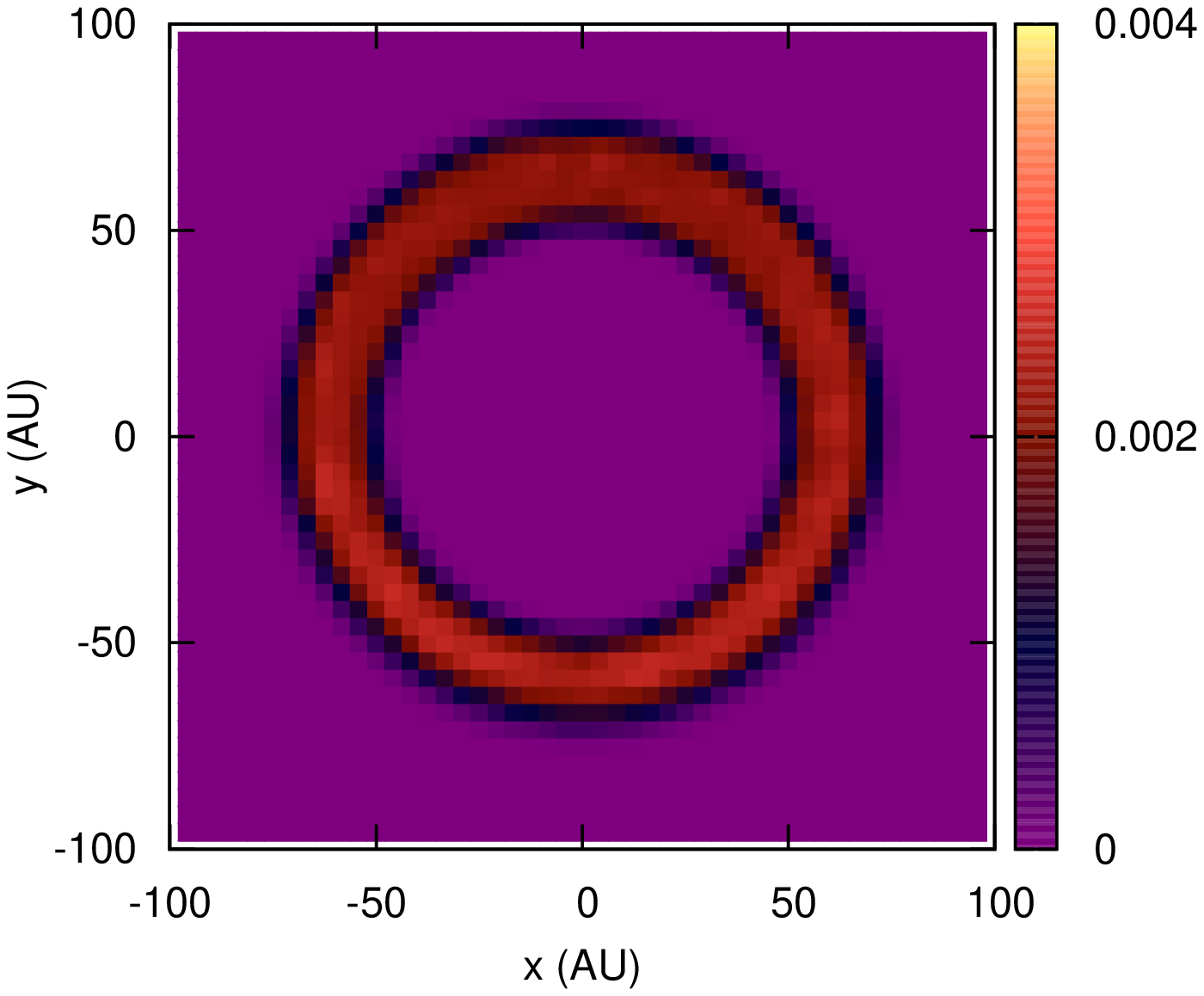}}\\
 \hskip -3 truecm
 \resizebox{120mm}{!}{\includegraphics[angle=-90]{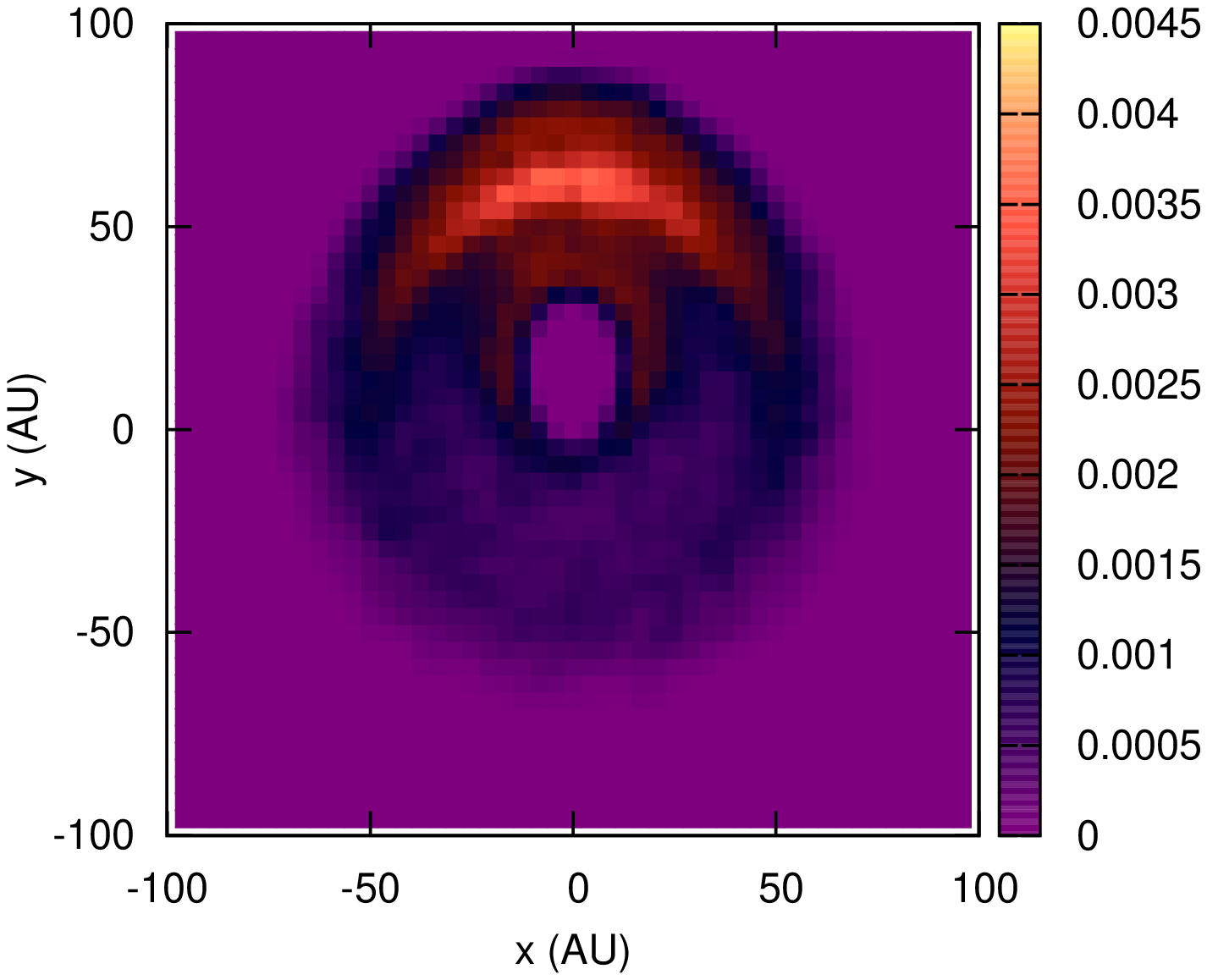}}
 \end{tabular}
 \caption{
2--D density distribution of 10 $\mu$m size dust particles when a steady 
state is reached. 
The top--left plot shows the case with 
optical depth $\tau = 10^{-3}$, the top--right plot that with $\tau = 10^{-4}$. 
The last and more perturbed case shown in the bottom plot corresponds
to an optical depth of $\tau = 10^{-6}$. This figure has to be compared
with Fig.\ref{f1}.}
\label{f3}
\end{center}
\end{figure*}

When we consider larger particles, a different balance between  
perturbative changes due to ISM and lifetime is reached. In addition,
the initial disk is expected to be less radially spread since the value 
of $\beta$ is smaller 
and 
the initial orbital elements of the grains are closer to those 
of the parent bodies. 
In Fig.\ref{f3} we show  density plots of a 
dusty disk made of 10 $\mu m$ size  particles ($\beta = 0.03$). 
If the optical depth is 
set to $\tau = 10^{-3}$ the disk does not show any difference 
from the case where ISM is neglected 
since the ISM perturbations do not have enough time to build up 
a consistent variation of eccentricity and perihelion. It
appears as a regular ring almost overlapping with the parent body
ring. The initial eccentricities are close to 0 
because of the small value of $\beta$ so that 
the disk appears circular and the 
overdense rings observed in Fig.\ref{f1},
top left plot, are not present.  
For a smaller optical depth, $\tau = 10^{-4}$, 
eccentricities as large as 
0.15 are pumped up by the ISM perturbations and the disk appears slightly eccentric. 
When finally $\tau = 10^{-6}$, the orbits of the grains reach a maximum 
eccentricity of $\sim$ 0.8 and a significant clump develops close to 
the apocenter of the orbits. In this case, the two--ellipse pattern,
observed for 1 $\mu$m in  Fig.\ref{f1} bottom--right plot, 
collapses into a single elliptical structure. The explanation for this different behaviour
resides in the lower initial eccentricity with which 10 $\mu$m 
are produced. This is about 0.03, small compared to the average value
of about 0.4 for 1 $\mu$m particles. By inspecting
Fig.\ref{f2} we notice that the analytical curve of the Stark problem 
for particles passing through low eccentricities gives values of 
perihelion very close to $270^o$ when the eccentricity approaches 1. 
All grains in the 10 $\mu$m case evolve from low eccentricity 
values  
and when their eccentricity is pumped up by the ISM perturbations, 
their pericenter is almost perfectly aligned to $270$. 
Exploiting again formula (33) of
\cite{bera}, when $ e = 0.03$ , we get an angular separation between the 
2 streams of particles, approaching $\omega=270^o$  for large eccentricities, of 
only $\sim 4^o$. As a consequence, in the density plot the 
two ellipses merge in a single one.

The behaviour shown in Fig.\ref{f3} confirms that the optical depth $\tau$ 
is the critical parameter in 
determining the amount of asymmetry produced by the ISM flux on a dust
disk even for larger sizes of the dust particles. When $\tau$ is large,
the effects of the ISM flux become negligible and the debris disk appears
symmetric. 

\section{The inclined case}

When the ISM flow is inclined of an angle $i_{ISM}$ 
respect to the parent body plane, the analytical model of the 
Stark problem predicts an increase of the grain orbital inclination  
and a reduction of the maximum value of the
eccentricity \citep{bera}. The 2--D model, described in the 
previous section, continues to be a good approximation only when $i_{ISM}$ is lower
than $10^o$. For larger values, the ISM perturbations lead to 
3--dimensional effects that alters the density distribution 
patterns observed in Fig.\ref{f1} and, in addition, cause significant
warping of the disk.  
However, if the optical depth $\tau$ is
of the order of $10^{-3}$,
the collisional disruption prevents the building up of 
significant inclination, as it does for the eccentricity,  
and the warping of the disk is negligible.

\begin{figure}
\hskip -2.1 truecm
 \resizebox{130mm}{!}{\includegraphics[angle=-90]{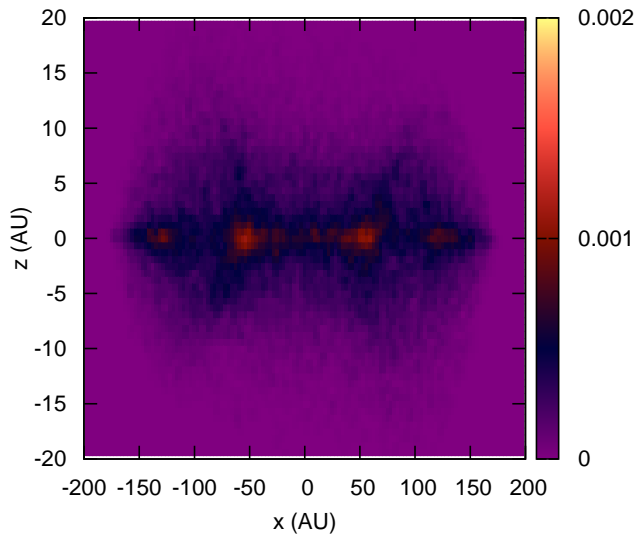}}\\
 \caption{Density distribution in the $(x--z)$ plane when 
$\tau = 10^{-3}$. The quick collisional disruption prevents large
out--of--plane structures to develop. Notice that the 
scale along the z--axis is 1/10 of the scale along the x--axis.
}
\label{f4}
\end{figure}

\begin{figure}
\begin{tabular}{c}
\hskip -2.1 truecm
 \resizebox{110mm}{!}{\includegraphics[angle=-90]{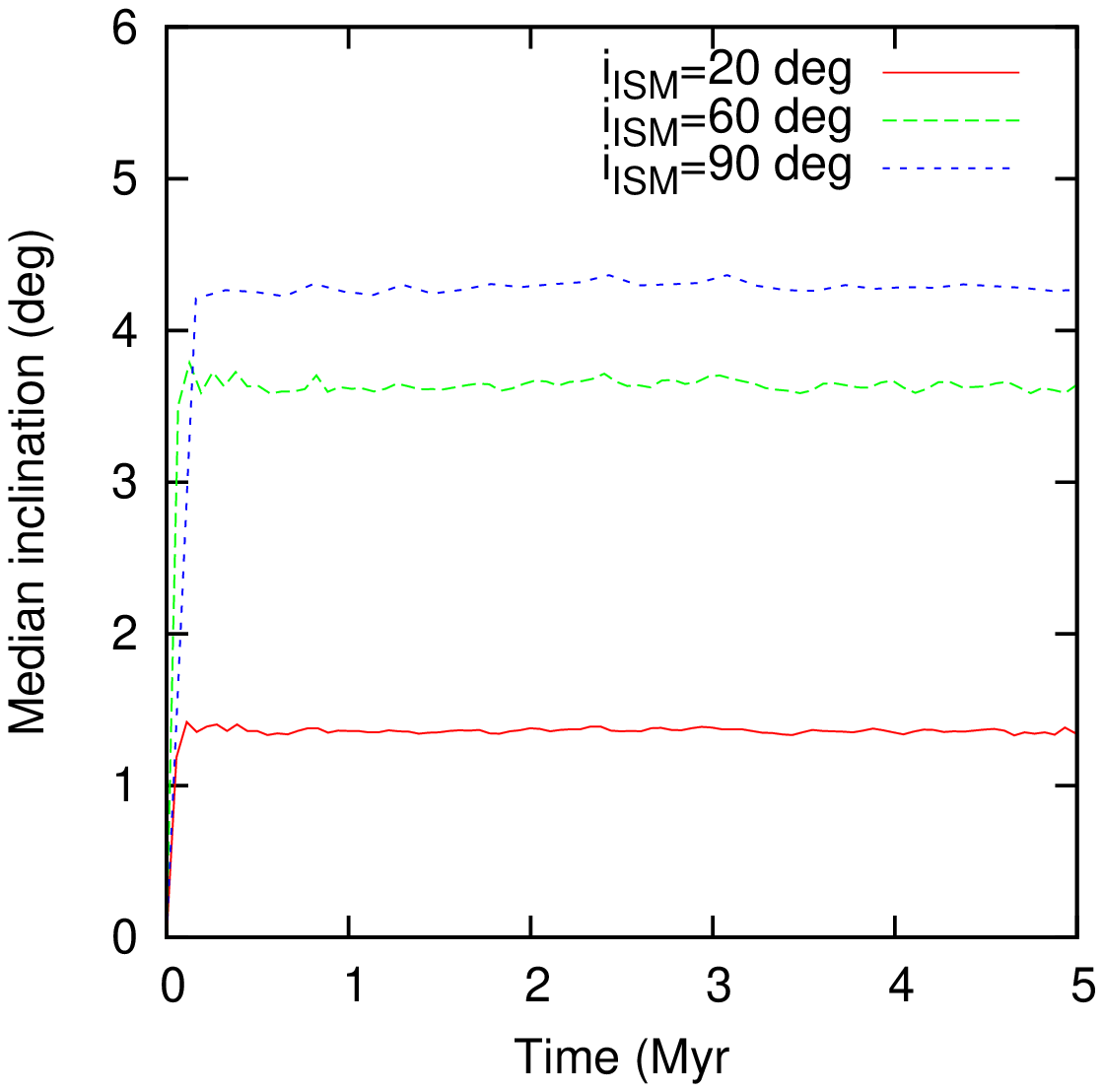}}\\
\hskip -2.1 truecm
 \resizebox{110mm}{!}{\includegraphics[angle=-90]{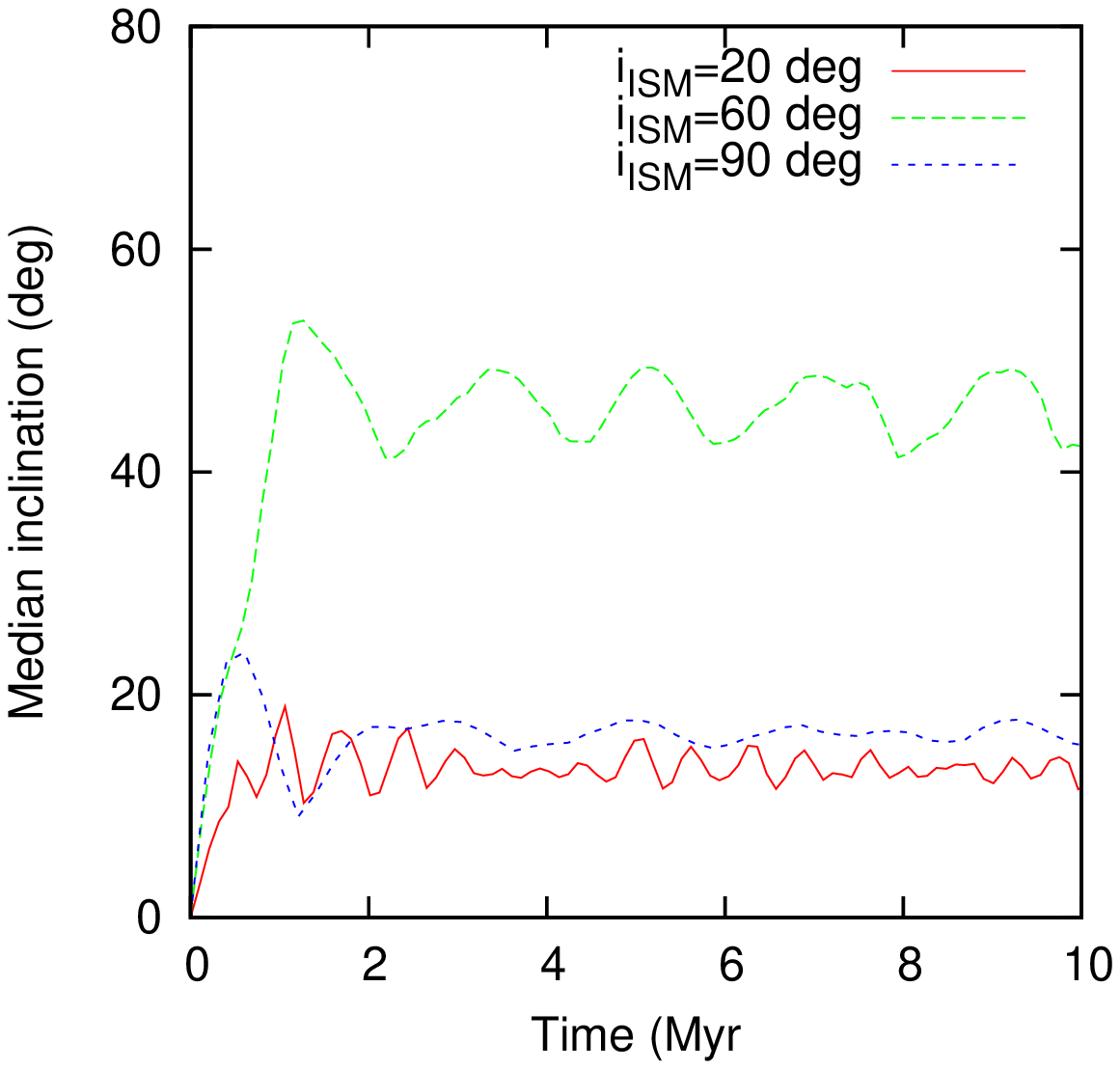}}
\end{tabular}
 \caption{Median inclination in a dust disk of 1 $1\mu$m size particles
for different values of $i_{ISM}$ and $\tau$. In the top plot 
$\tau = 10^{-3}$ while in the bottom one it is $\tau = 10^{-6}$.  
}
\label{f5}
\end{figure}

This is shown in  Fig.\ref{f4} where the outcome of a simulation 
with $i_{ISM}=60^o$ and $\tau = 10^{-3}$ is illustrated as 
a density plot in the $(x-z)$ plane. Only 
small out--of--plane features can be detected in the figure which
cannot be interpreted neither as warping or clumping due to their limited 
extension along the z--axis (the scale along the z--axis is 1/10 
of the scale along the x--axis). 
A large optical depth 
prevents then a disk to develop structures not only in the parent
body plane but also out--of--plane. This finding is confirmed 
for any value of $i_{ISM}$.  
In Fig.\ref{f5} the median inclination of the grains in
the disk 
is plotted for different values of $i_{ISM}$  with 
$\tau = 10^{-3}$ (top plot).  The value of the median
of the particle inclinations is always smaller than $5^o$ since the 
Stark cycle is quickly interrupted by a collision. This confirms that 
if the disk is optically thick, the ISM does not have time to 
build up observable signatures in the density distribution of debris disks. 
In addition, the particles do not have the time to drift inside since 
they are destroyed before PR drag and the ISM drag force them 
to migrate inside. If a gap is present in the inside regions of the
disk this will not be filled up by the grain migration.

\begin{figure}
\hskip -2.1 truecm
\subfigure{
 \resizebox{110mm}{!}{\includegraphics[angle=-90]{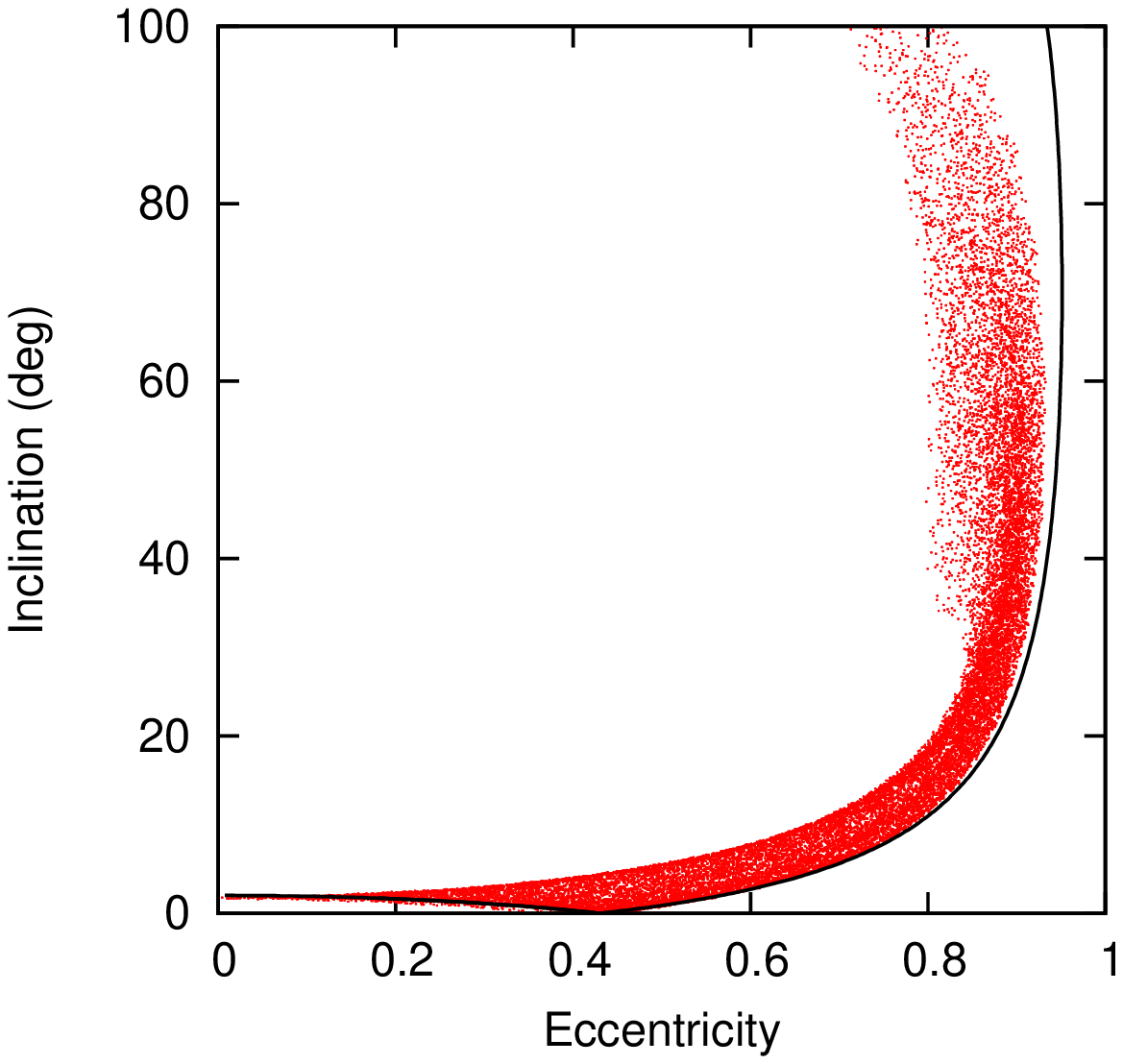}}} \\
\vskip -1.0 truecm 
\hskip -2.1 truecm
\subfigure{
 \resizebox{110mm}{!}{\includegraphics[angle=-90]{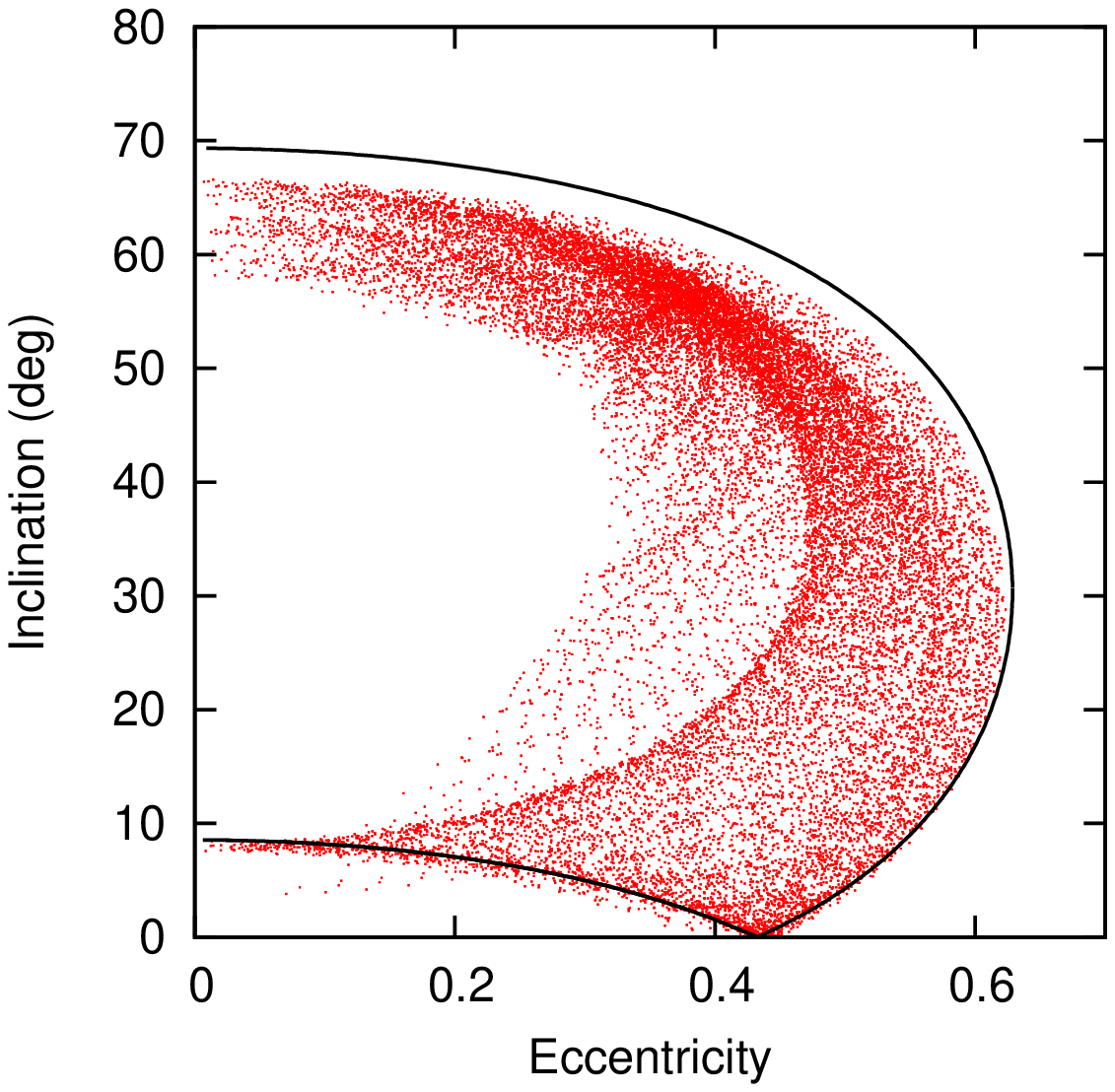}}}\\
\vskip -1.0 truecm 
\hskip -2.1 truecm
\subfigure{
 \resizebox{110mm}{!}{\includegraphics[angle=-90]{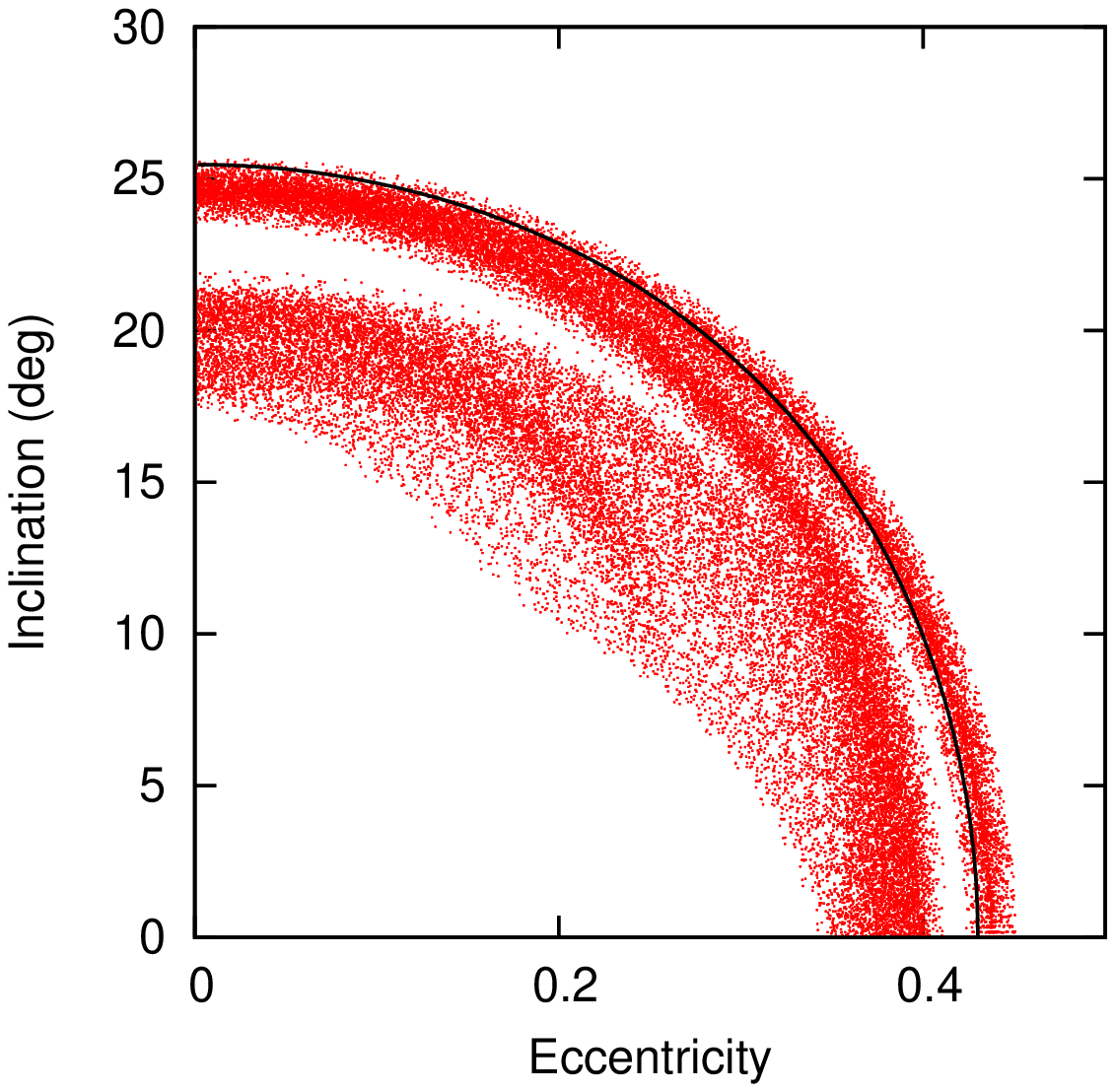}}}
 \caption{Correlation between the eccentricity and inclination for 
dust particles in a disk with different values of $i_{ISM}$. 
In the top plot $i_{ISM} = 20^o$, in the middle $i_{ISM} = 60^o$ and
in the bottom $i_{ISM} = 90^o$. The continuous lines are the  theoretical 
predictions derived from the analytical theory of the Stark problem. 
}
\label{f6}
\end{figure}

Significantly different is the situation when 
$\tau$ is as low as $10^{-6}$ (Fig.\ref{f5} bottom plot): particles in the disk have
time to move along the Stark cycle. This is not interrupted by 
collisions since the collisional lifetime is longer than 
$T_{stark}$ and the particles can build up large orbital 
eccentricities and inclinations.
In this scenario, the only mechanism able to halt the Stark 
cycling is the fast inward drift due to the large eccentricity. 
The particles migrate inside driven by the PR and ISM drag 
forces  ending either onto the star or sublimating or 
impacting onto a planet.  This sink
mechanism is more effective for smaller values of $i_{ISM}$ since
the maximum eccentricity $e_{max}$ that can be achieved by the particles 
decreases for increasing $i_{ISM}$, according to the Stark general problem. 
In  Fig.\ref{f6} we show the evolution of the eccentricity and 
inclination for increasing values of $i_{ISM}$. Their values
are strongly correlated, as predicted by the theory of 
\cite{bera} (the theoretical curves are plotted as continuous lines in the figures). 

Two interesting features come out from Fig.\ref{f6}.
First of all the drag forces tend to reduce the 
eccentricity detaching the numerical data from the theoretical  curve 
even if the difference is not very marked. 
In addition, when the maximum eccentricity 
is lower, for higher $i_{ISM}$, the  particles move farther within the Stark cycle 
because their drift time is longer. When $i_{ISM} = 90^o$ the particles 
complete a full cycle and some start a new one, even if they do not complete 
more than two cycles before drifting inside 20 AU. 

\begin{figure*}
 \hskip -3 truecm
\subfigure{ 
\resizebox{120mm}{!}{\includegraphics[angle=-90]{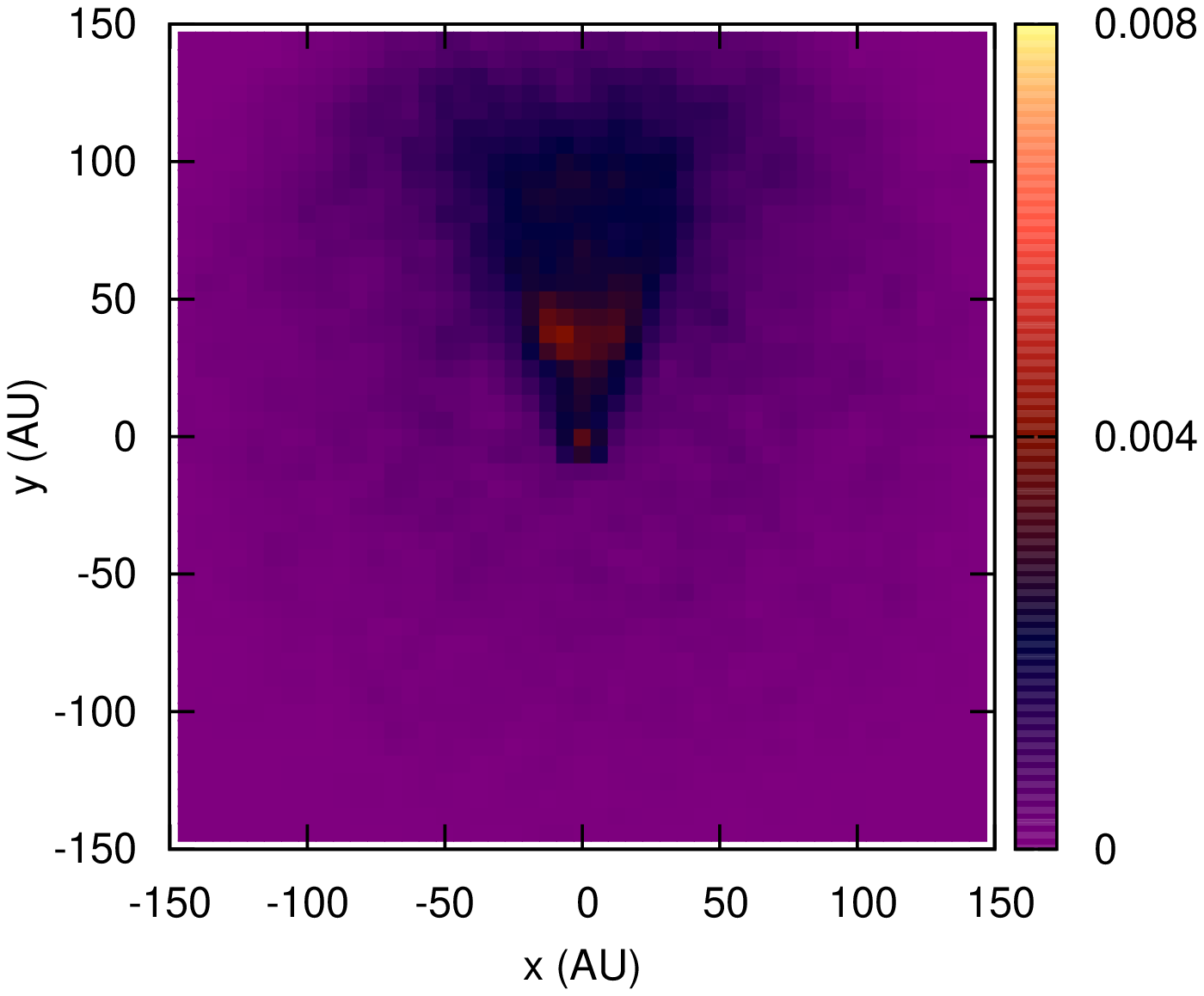}} \hskip -4 truecm
 \resizebox{120mm}{!}{\includegraphics[angle=-90]{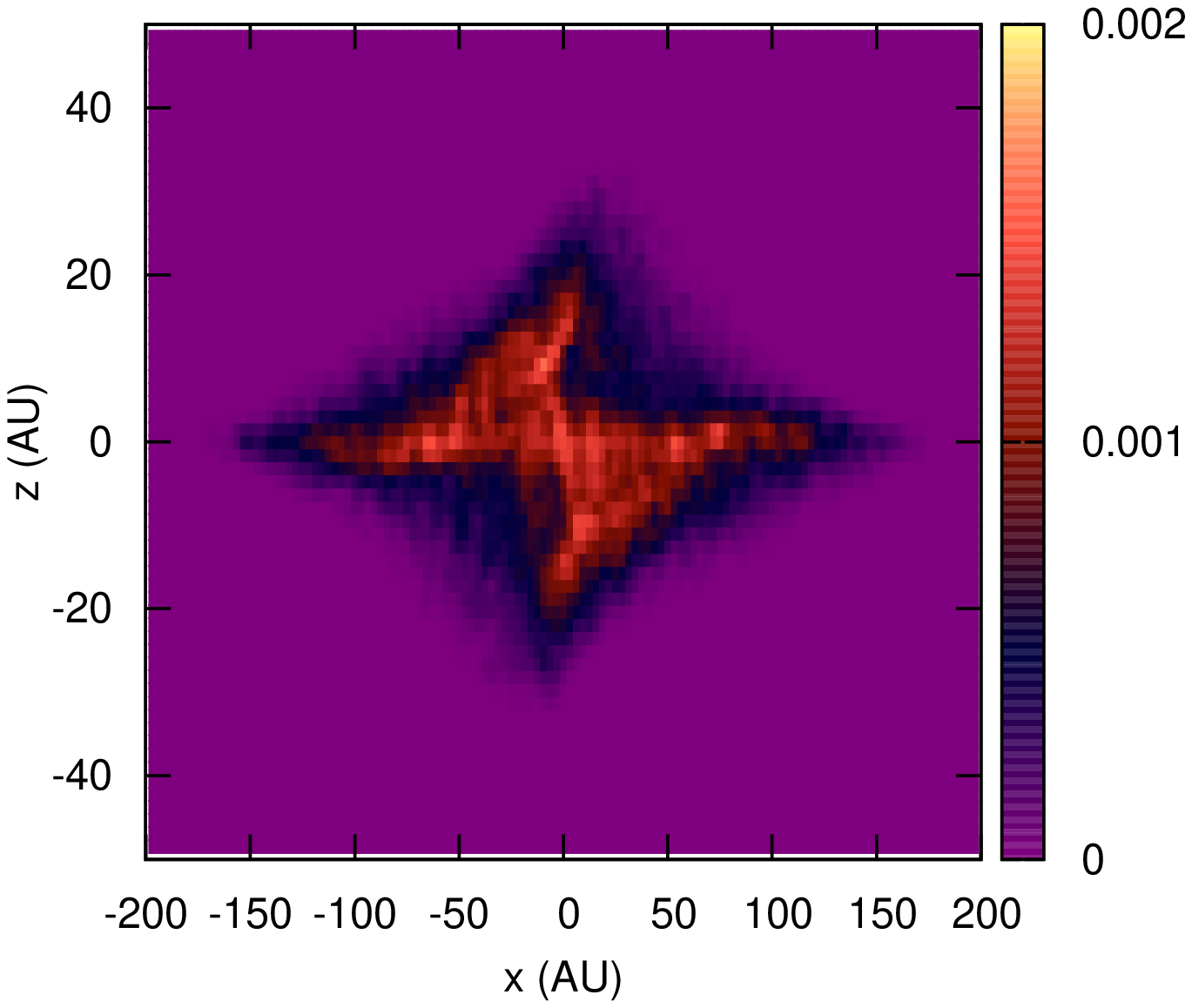}}}\\
 \vskip -2 truecm
 \hskip -3 truecm
\subfigure{
 \resizebox{120mm}{!}{\includegraphics[angle=-90]{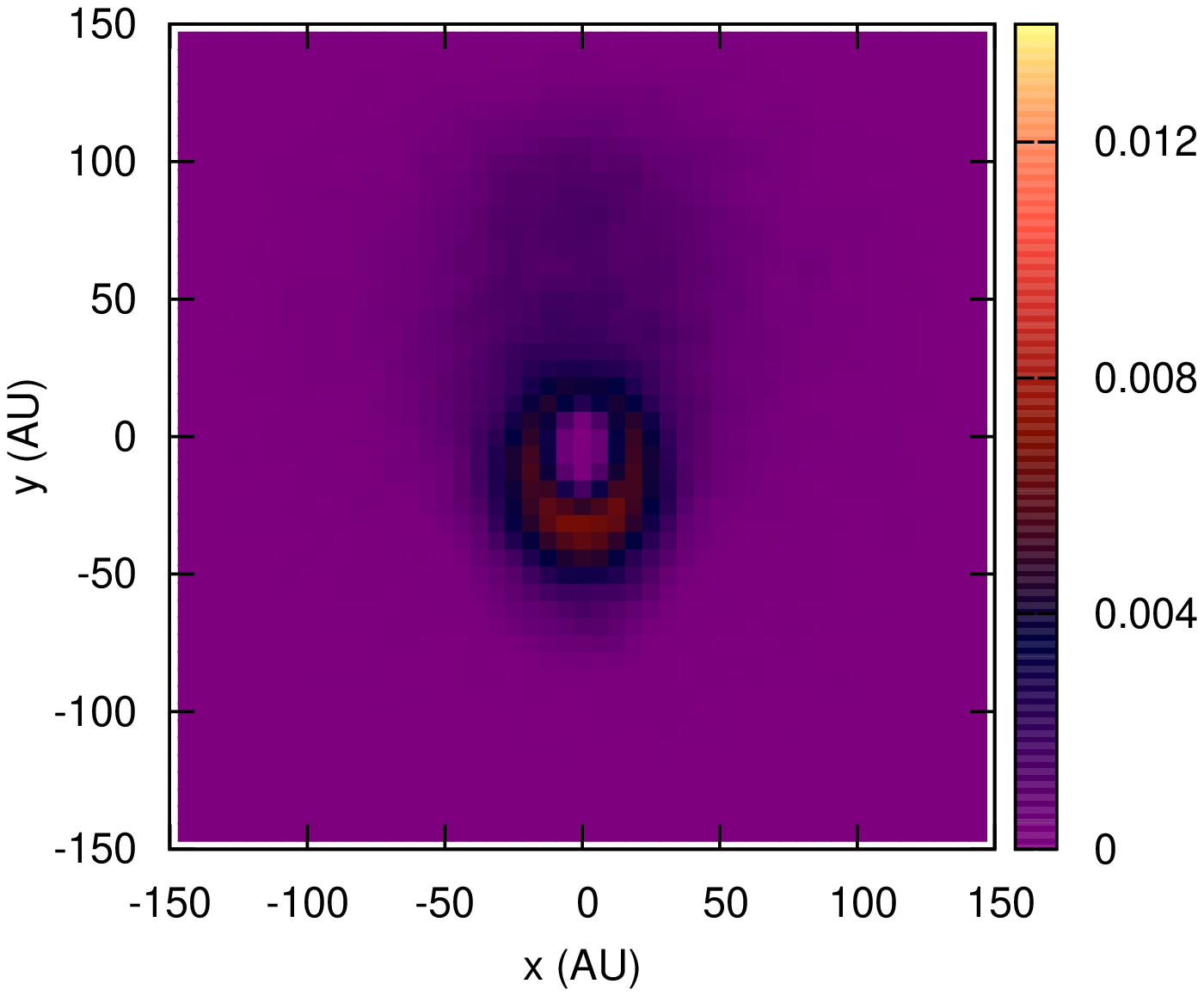}} \hskip -4 truecm
 \resizebox{120mm}{!}{\includegraphics[angle=-90]{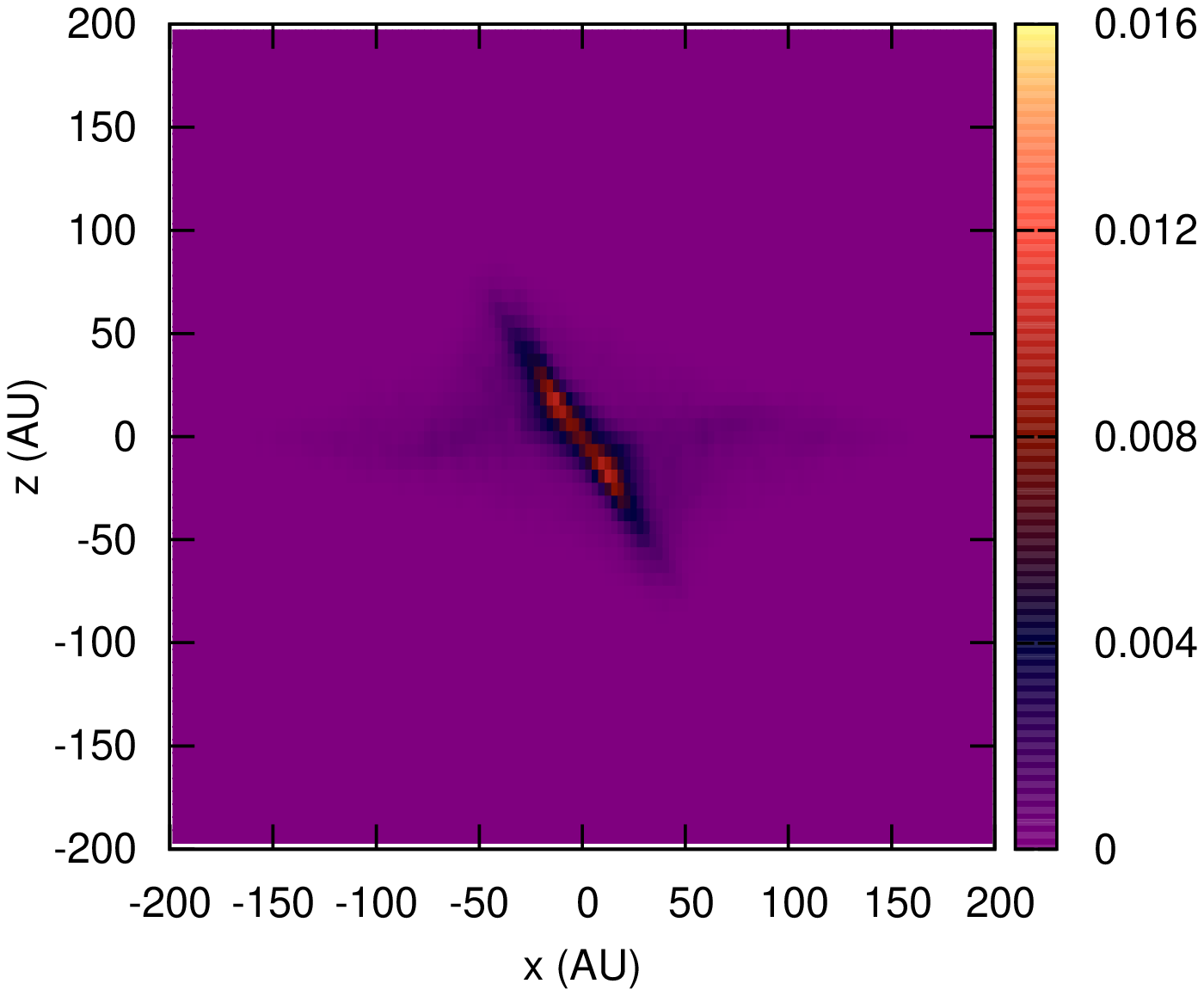}}}\\
 \vskip -2 truecm
 \hskip -3 truecm
\subfigure{
 \resizebox{120mm}{!}{\includegraphics[angle=-90]{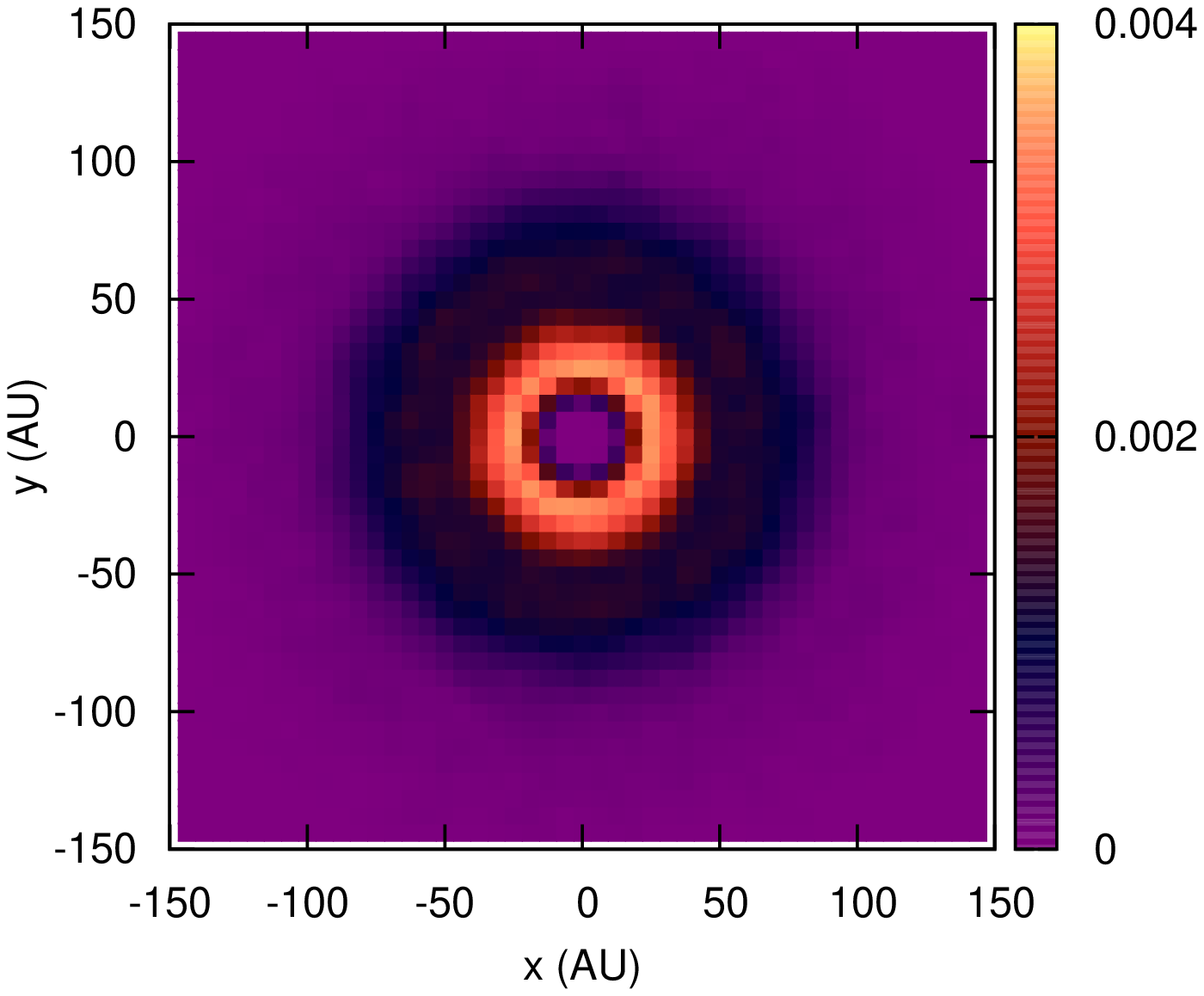}} \hskip -4 truecm
 \resizebox{120mm}{!}{\includegraphics[angle=-90]{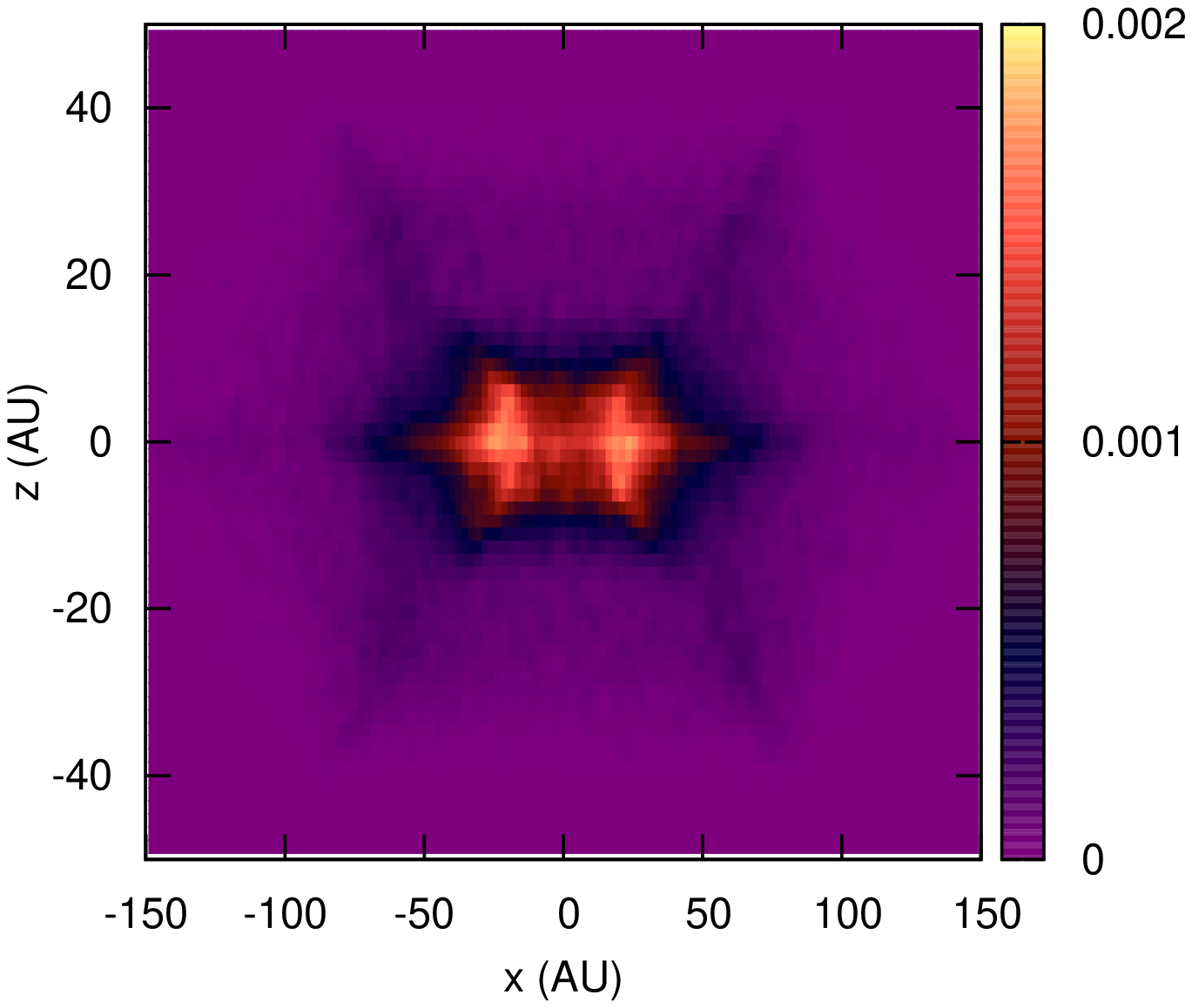}}}\\
 \caption{Density distributions of 1 $\mu$m size dust particles in the 
$(x--y)$ and $(x--z)$ plane for different values of $i_{ISM}$. The 
top plots refer to the case $i_{ISM}=20^o$, the middle plots to 
$i_{ISM}=60^o$ and the lower plots to $i_{ISM}=90^o$,
respectively.}
\label{f7}
\end{figure*}

Additional features related to the dynamics of the dust particles under 
the action of the ISM flow and PR drag can be seen in the density plots
shown in Fig.\ref{f7}. In the parent body plane the density distribution
appears very asymmetric. In the case where $i_{ISM} = 20^o$ 
(Fig.\ref{f7} top plots) it is still 
possible to recognize the two elliptical structures produced by the 
evolution in the $(e,\omega)$ plane but their shape is blurred due 
to the inclination distribution. However, when $i_{ISM} = 60^o$ the 
double--elliptical structure has fully disappeared and the inclination distribution 
determines the density distribution also in the parent body plane. 
As it can be argued from 
Fig.\ref{f6} middle plots, the orbital inclination is concentrated 
around $i \sim 50^o$ and, in addition, when the inclination is higher,
the node longitude is clustered around $\Omega = \pi /2$. This 
leads to an eccentric disk structure that is inclined respect to the original 
orbital plane of the parent body of approximately $50^o$. Finally, the 
case $i_{ISM} = 90^o$ (Fig.\ref{f6} bottom plots) is the most symmetric case. 
The circular overdense 
shape of the disk close to the star is where the inclined orbits 
cross the parent body plane. The dust particles have a statistical uniform 
distribution of the nodes for any value of inclination and, as a consequence, 
a high density region is formed in correspondence to the intersection 
between the inclined orbits and the initial parent body plane. 

\begin{figure}
 \hskip -3 truecm
\subfigure{
 \resizebox{110mm}{!}{\includegraphics[angle=-90]{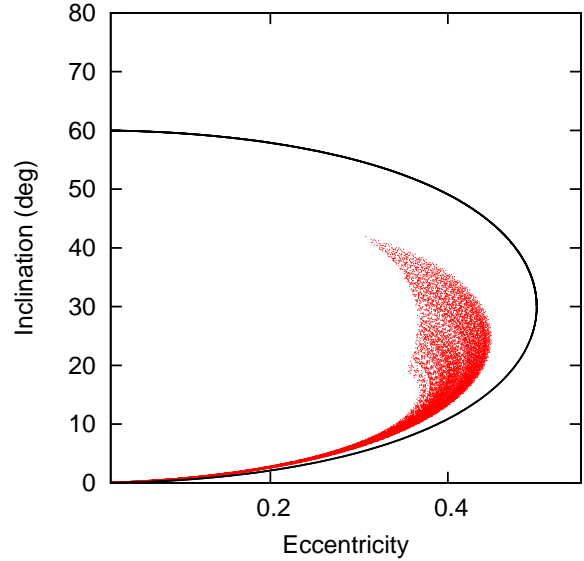}}}\\
 \caption{Eccentricity--inclination distribution for 10 $\mu$m size
dust particles in a disk with $i_{ISM} = 60^o$. 
}
\label{f8}
\end{figure}

\begin{figure*}
 \hskip -3 truecm
\subfigure{
 \resizebox{120mm}{!}{\includegraphics[angle=-90]{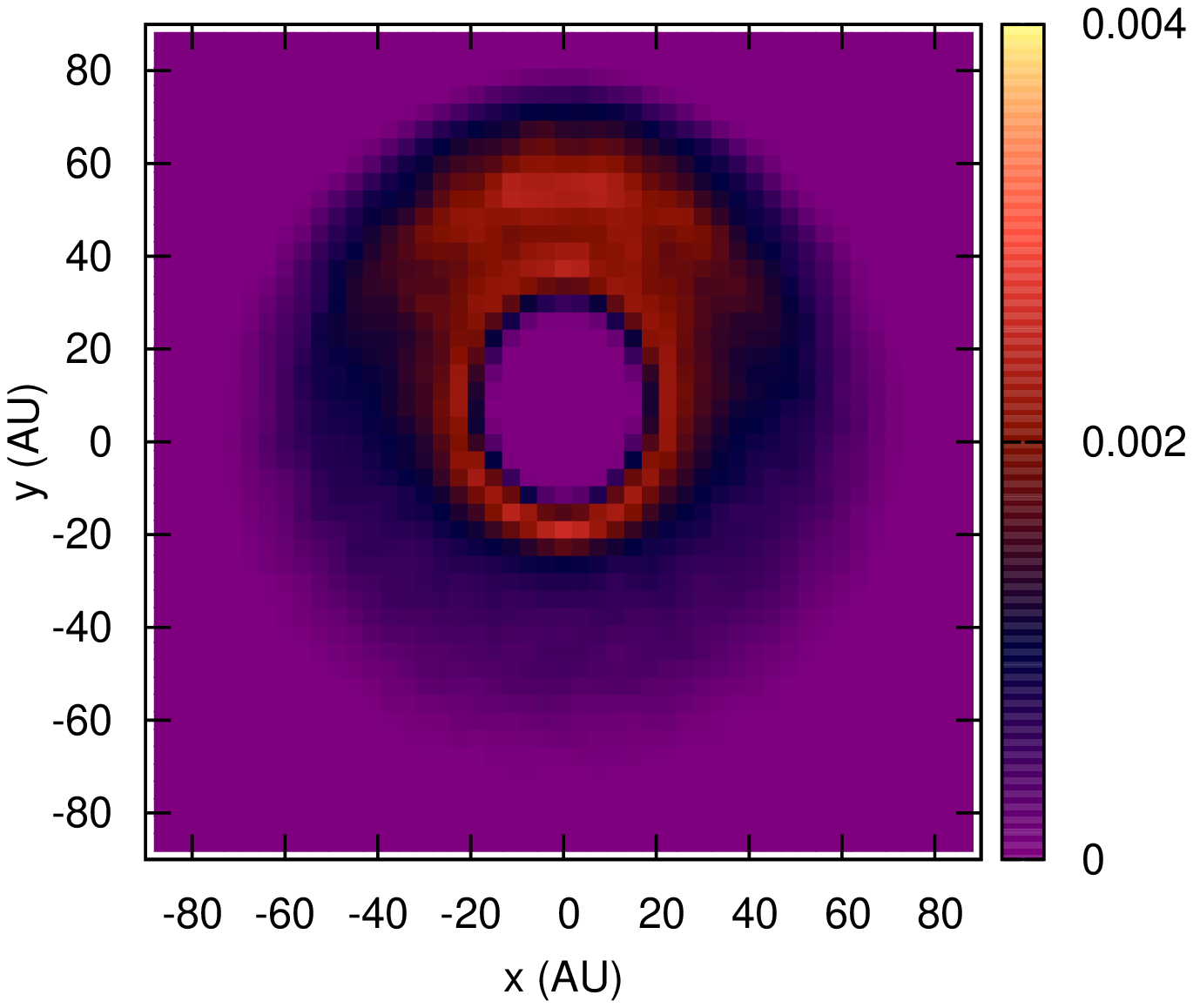}} \hskip -4 truecm
 \resizebox{120mm}{!}{\includegraphics[angle=-90]{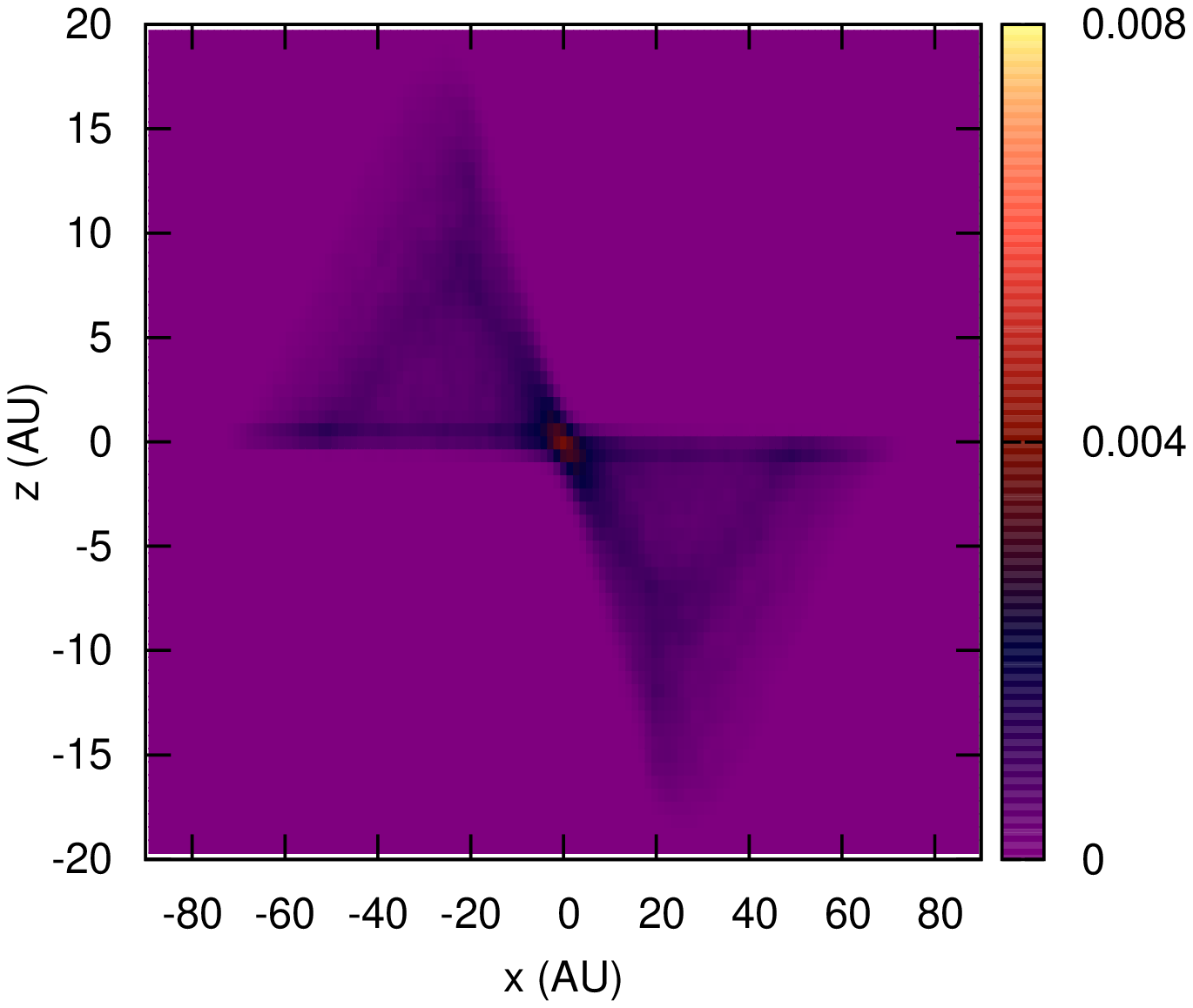}}}\\
 \caption{Density distributions of 10 $\mu$m size dust particles in the 
$(x--y)$ and $(x--z)$ plane for $i_{ISM}= 60^o$. These distributions 
can be compared to those for 1 $\mu$m shown in Fig.\ref{f6} middle
plots.
}
\label{f9}
\end{figure*}

For larger dust grains (10$\mu$m) and low optical depth ($\tau = 10^{-6}$), 
the dynamical scenario is slightly different. The Stark period is of the 
order of 23 Myr, 10 times longer than that of 1$\mu$m particles, but the 
drift rate has not decreased by the same amount. As a consequence, the 
grains evolve less into the Stark cycle and migrate inside before 
reaching large values of inclination. This is shown in Fig.\ref{f8} 
where the eccentricity and inclination of the dust, when the disk is 
in a steady state, are illustrated. The initial eccentricity is close to 
0 since $\beta$ is small and this leads to a different Stark cycle compared 
to that in Fig.\ref{f6} middle
plots. The particles slowly evolve towards larger inclination values 
but they plunge into the inner regions of the disk and are lost before they 
can develop significant inclinations. The disk is then 
expected to be less perturbed compared to the case for 1 $\mu$m size particles.
This is confirmed by Fig.\ref{f9} where the spatial density distribution 
is illustrated and can be compared to that shown in Fig.\ref{f7} (middle plot)
for 1 $\mu$m size particles. 

\section{Conclusions}

We have shown in this paper, by numerical modeling the 
evolution of debris disks 
under the effects of solar radiation pressure, PR drag 
and ISM flux, that  
for typical values of optical depth $\tau \sim 10^{-3}$ the signatures of  
the ISM wind on grains with radius in the range $1-10$ $\mu$m,
just above the cut-off size
imposed by radiation pressure,
are almost negligible. The amount of perturbations
due to 
the interaction of dust particles with the local ISM
neutral atom flow are strongly reduced when the grain lifetime 
is short as in disks with large values of $\tau$. In this scenario, the 
presence of asymmetries must be ascribed to different mechanisms 
like the presence of massive bodies within the disk.  The 
neutral ISM fails in producing either warping or clumping 
in such disks. 
 
Different is the scenario when the optical depth is small. 
We have shown that for $\tau \sim 10^{-6}$  
significant 
asymmetries appear on the density profile of the disk 
both in the parent bodies plane and out--of--plane. 
Interesting is a double--elliptical pattern that 
develops in the parent bodies plane when the ISM flow 
is almost coplanar to it.  
The disk structures are  caused by 
the dynamical evolution of the orbital elements 
of the
dust grains. 
The way in which these orbital parameters evolve can be
in part predicted on the basis of the general Stark model 
that helps in  
understanding the periodic nature 
of eccentricity, inclination and the dynamically related 
angles perihelion argument and nodal longitude  
\citep{bera,pasto}. In addition, the 
semimajor axis has a fast inward drift due to the 
combination 
of PR drag and interaction with the ISM. 
The two drag forces, in particular that related
to the ISM flow, are strong because of 
the large ecceentricity of the grains.
The particles quickly migrate towards the 
star 
where they can either sublimate or 
be destroyed by collisions with planets or the star itself. This is an
additional powerful sink mechanism for debris disks with small 
optical depth. 
On the contrary, if the disk has a large optical depth, collisional disruption 
acts on a much shorter timescale and it prevents a significant 
inward migration. The disk shape in this case would be  
mostly due to the initial value of $\beta$ which sets the 
initial orbital element distribution of the grains.  
The debris disk would then extend outwards respect to the 
parent body ring in spite of the strong ISM and radiation drag force. 
This additional different dynamical behaviour of the dust grains,
which again depends on the optical depth, sets a nice 
correlation between the value of $\tau$ and location of the disk
respect to the parent body ring. 
Disks with low values 
of $\tau$ would expand inward respect to the parent body ring
down to the sublimation region or where planets are orbiting, 
while disks with large values of $\tau$ would extend mostly outwards.

Our study suggests that  observational 
data of debris disks need care to be interpreted. Potential asymmetries 
identified in the density distribution 
may be ascribed to interactions with the local 
flux of ISM neutral atoms surrounding its parent star
only if the disk has a low optical depth. Otherwise,  
alternative explanations, like the presence of planets, must be investigated. 

\section*{Acknowledgments}

We thank Mikhail Belyaev for his useful comments and suggestions while 
acting as referee of the paper.

\end{document}